\documentclass{osa-article}

\journal{oe}

\usepackage{subfigure}
   \newcommand{\blue}[1]{{\color{blue}#1}}
\usepackage{ulem}
\articletype{Research Article}

\begin{document}

\title{Inverse design method for periodic and aperiodic metasurfaces based on the adjoint-method: metalens with random-like distributed nano-rods}
\author{Kofi Edee \authormark{1,*},  Mauro Antezza \authormark{2,3},  Brahim Guizal\authormark{2}}
\address{\authormark{1}Universit\'{e} Clermont Auvergne,
Institut Pascal, BP 10448, F-63000 Clermont-Ferrand, France, CNRS, UMR 6602, Institut Pascal, F-63177 Aubi\`{e}re, France}
\address{\authormark{2}Laboratoire Charles Coulomb (L2C), UMR 5221 CNRS-Universit\'{e} de Montpellier, F-34095 Montpellier, France}
\address{\authormark{3}Institut Universitaire de France, 1 rue Descartes, Paris Cedex 05 F-75231, France}
%
\email{\authormark{*}kofi.edee@uca.fr}



\begin{abstract}
The classical adjoint-based topology optimization (TO) method, based on the use of a random continuous dielectric function as an adjoint variable distribution, is known to be one of the most efficient optimization methods that enable the design of optical devices with outstanding performances. However, the strategy for selecting the optimal solution requires a very fine pixelation of the permittivity function of the profile under optimization.  
Typically, at least $2^8$ pixels are needed while optimizing a one wavelength wide 1D metagrating. This makes it very difficult to extend TO methods to large-scale optimization problems. In this paper, we introduce a new concept of adjoint-based topology optimization that enables fast and efficient geometry based design of both periodic and aperiodic metasurfaces. The structures are built from nano-rods whose widths and positions are to be adjusted. Our new approach requires a very low number of design parameters, thus leading to a drastic reduction in the computational time: about an order of magnitude. 
Hence, this concept makes it possible to address the optimization of large-scale structures in record time. As a proof-of-concept we apply this method to the design of (i) a periodic metagrating, optimized to have a specific response into a particular direction, and (ii) a dielectric metalens (aperiodic metasurface), enabling a high energy focusing into a well-defined focal spot. Besides, a common strategy practiced in the design of large structures consists in optimizing sub-parts of the desired structure before assembling them under some phases constraints to yield the final device. One drawback of such a method is that it generally leads to symmetric structures and thus ignores a huge set of other, possibly better, solutions in the optimization space. In our approach, we optimize the full device in a single play, without enforcing any specific phase conditions, and demonstrate that the proposed method systematically provides unexpected asymmetric metalenses with highly non regular nano-rods distributions. To the best of our knowledge, this kind of metalenses with a random-like distributed nano-rods has never been reported before in the literature.  
\end{abstract}

\section{Introduction}
In photonics, the design of metasurfaces can be thought of as a methodology allowing to shape or structure the surfaces of devices, at a subwavelength scale, in order to have specific responses to particular excitations. In recent years, the use of inverse design methods, and in particular those based on forward and adjoint computations, has proven to be one of the most efficient in practice.  
Considering a given target solution (e.g. a given electromagnetic field distribution in a given space), the aim of inverse design methods is to find a physical configuration associated with this solution. Mathematically speaking, considering a set of equations parameterized by a sequence of variables commonly called design parameters, one often wishes to optimize a quantity, usually called figure of merit (FOM), based on these parameters and on the solutions of the equations under consideration. For example, in photonics, given the deflected power in a specific direction by a 1D structured slab, one may be interested in finding a sub-structure of that slab made of a given number of subwavelength rods that generates this solution \cite{Sell1, Yang1, Sell2, Yang, EWang, Kofi_TO, Kofi_TO2}. Or in the case of metalenses design, one can be interested in finding a structure enabling to focus a maximum power flow into a particular focal point \cite{Shrestha, SWang, Phan, Chung}. In practice, knowing the FOM is extremely useful, and even more important is the knowledge of its gradient $g$. This latter measures the sensitivity of a desired response to the design  parameters, and thus, indicates a search direction that enables the increase of the FOM. However, the number of the design parameters is often very large resulting in very expensive computational times when evaluating the gradient. This difficulty can be circumvented thanks to adjoint methods that allow an evaluation of $g$ with a cost independent of the number of the design parameters. Recently, some adjoint-based-topology optimization (TO) methods, based on the use of continuous functions with respect to the design parameters, and providing impressive high performance results, have been proposed \cite{Lu, Lalau, Hughes,Molesky,Frandsen, Borel, Piggot, Xiao, Lin}. 
In these approaches, the starting point, a continuous profile describing the electromagnetic properties of the structure, is numerically approximated by a sequence of pixels; and, at each iteration $t$, the gradient of the figure of merit $g$ is computed in each voxel (volume element) of the design area.  This gradient is then used to modify the design variable values at each voxel location in order to increase the FOM. The algorithm is performed iteratively by pushing the continuous profile towards a  discrete one. And while converging to a discrete profile, filtering (blurring) and projection (binarization) schemes are periodically  applied. 
To achieve optimized structures with high performances, a very high number of pixels is required in order to fine-tune the geometrical profile during the iterations. Typically, at least  $2^8$ pixels are needed for the optimization of a one dimensional metagrating of few wavelengths scale and this number dramatically increases with the size of the device. Hence this TO method is unsuitable for the design of large-scale structures such as aperiodic metasurfaces. \\

In this paper, instead of modifying continuously the value of the permittivity function at each voxel-location in the design domain, we propose a faster and more efficient method. This will be done by acting directly on the widths of the metasurface constituents, namely a sequence of subwavelength rods and air-gaps in the case of a one dimensional structure (Figure \ref{TO_DIRECT_ADJOINT_TO_lens2}). The algorithm is then performed by continuously modifying their widths and their locations. Our new design variables become the nano-rods and air-gaps widths. Contrary to topology optimization (TO) methods proposed and used in \cite{Sell1,Yang1, Sell2, Yang, Phan,EWang, Kofi_TO,Kofi_TO2}, our optimization starts with a piece-wise permittivity function describing a discrete distribution of subwavelength (rods, air-gaps) elements.\\ 

In section 2, we present the mathematical formulation of the proposed concept. In section 3, we first demonstrate its ability to design a 1D metagrating capable of maximizing light diffraction into a given deflection angle. Later, we extend the method to aperiodic metasurfaces which we apply to a dielectric metalens. In this example, we show that our optimization process yields an asymmetric metalens enabling on-axis focusing of an incident plane wave under normal incidence. To the best of our knowledge, this kind of metalens with a random-like distributed nano-rods has never been reported before in the literature.   
\section{Methods}
The concept proposed in this paper, and schematized in Figure \ref{TO_ALGO},
is different from the classical adjoint-based TO approach \cite{Lu, Lalau, Hughes, Molesky,Frandsen, Borel, Piggot, Xiao, Lin, Sell1, Yang1, Sell2, Yang, Phan, EWang, Kofi_TO, Kofi_TO2}, which replaces an absolutely continuous permittivity function by a  discrete distribution through the tedious processes of filtering and binarization. This leads to very long computational times when the number of design parameters is large. In the method we propose, only a small number of key design parameters is kept yielding a far less expensive electromagnetic computations. 
Like most of gradient-based methods, our concept can be viewed as a concatenation of three phases that are performed gradually and iteratively: 
\begin{itemize}
\item Initialization
\item Evaluation of the fitness of the design parameters 
\item Best current pattern update
\end{itemize}
In the following subsections, we detail the principles of each of these steps. 
\subsection{Initialization}
\begin{figure}[htb!]
 \centering
   \subfigure [\label{TO_ALGO} ]
        {\includegraphics[width=.8\textwidth]{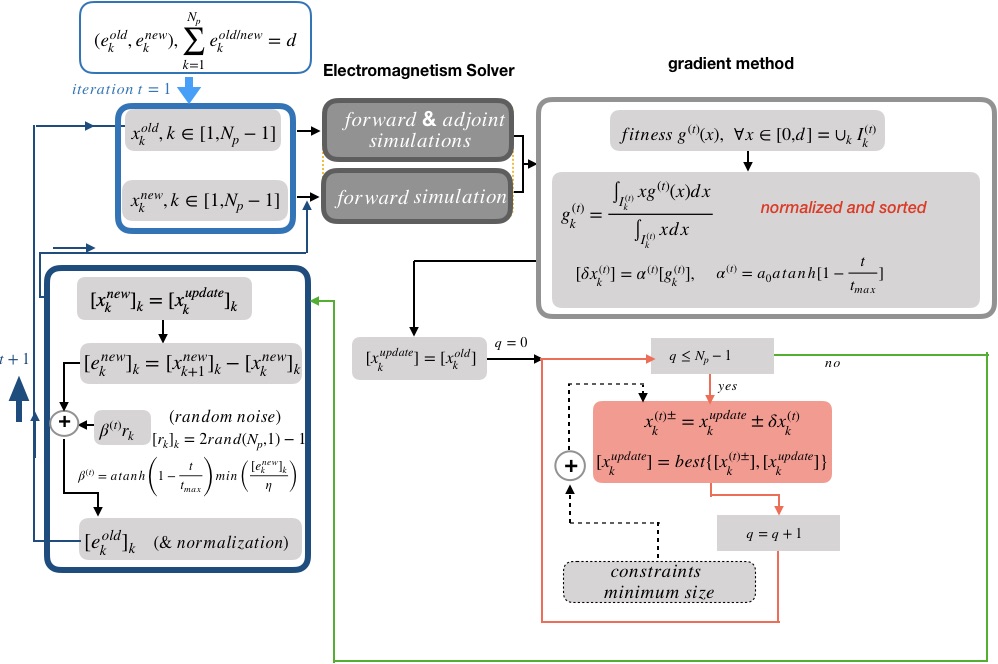}}
\centering
   \subfigure [\label{TO_deltaxi} ]
        {\includegraphics[width=.5\textwidth]{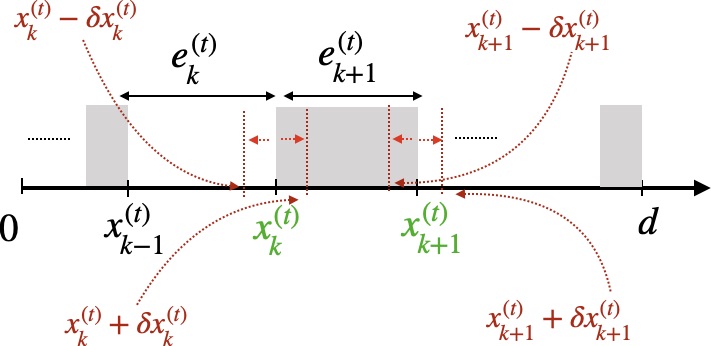}}
\caption{\label{TO_ALGO_TO_deltaxi}
Fig. \ref{TO_ALGO}: Flowchart of the proposed topology optimization method. In order to highlight the versatility and the flexibility  of the algorithm, only a simplified  flowchart is presented. Fig. \ref{TO_deltaxi}: Example of border-location variables $x_k$ update. In the proposed TO process, the increment increasing favorably the FOM is kept as the best current optimal profile.}
\end{figure}

In the initialization phase, the algorithm starts  by  generating two groups of $N_p$-tuple random  variables $(e_k^{old})$ and $(e_k^{new})$ inside a range $[e_{min},e_{max}]$. These random variables are associated with the following set  of nano-rods and air-gaps widths:  
\begin{equation}\label{width1}
    e_k^{old/new}=(e_{max}-e_{min})r^{old/new}_k+e_{min}, 
\end{equation}
where $r_k^{old/new}$ are a sequence of random variables in  $[0,1]^{N_p}$. These sequences of widths must satisfy the condition : $d=\sum_{k=1}^{N_p} e_k^{old/new}$ where $d$ is the size of the structure (or its period, in the case of metagratings). This is why we normalize them as follows: 
\begin{equation}\label{width2}
    e_k^{old/new}\rightarrow \dfrac{d\times e_q^{old/new}}{\sum_{q=1}^{N_p} e_k^{old/new}}  
\end{equation}


Such variables choice, \textsl{i.e.}, the widths as adjoint variables, restricts the design space by setting the locations of both the rods and the air-gaps. To circumvent this drawback and allow more degrees of freedom, we introduce a new sequence of variables $x_k$, termed border-location-variables and defined from $e_k$ by : 
\begin{equation}\label{x_interface}
    x_k=x_{k-1}+e_k,\quad x_0=0, \quad k\in[1,N_p-1].
\end{equation}
These new variables, displayed on Figure \ref{TO_deltaxi}, enable an easier and fine-tuning of both the widths and the locations of the rods, within the optimization process.

\subsection{Fitness evaluation}
At the $t^{th}$ iteration, the adjoint based method \cite{ Hughes, Piggot, Sell1}, is used to compute, in a single run, the gradient $g^{(t)}(x_i)$ of the objective function at all nodes $x_i$ of the design area. 
The key point of the gradient function computation, already discussed in \cite{Kofi_TO}, is to consider that fictitious currents  are induced when the system under consideration transits from a state called $old$ to a state called $new$. 
This transition may be due to an evolution of the geometrical and/or the physical parameters of the system. A more detailed discussion on the computation of the gradient of the FOM, applied here, is provided in references \cite{Kofi_TO, Kofi_TO2}.

\begin{figure}[htb!]
 \centering
    \subfigure [\label{TO_DIRECT_ADJOINT} ]
         {\includegraphics[width=.6\textwidth]{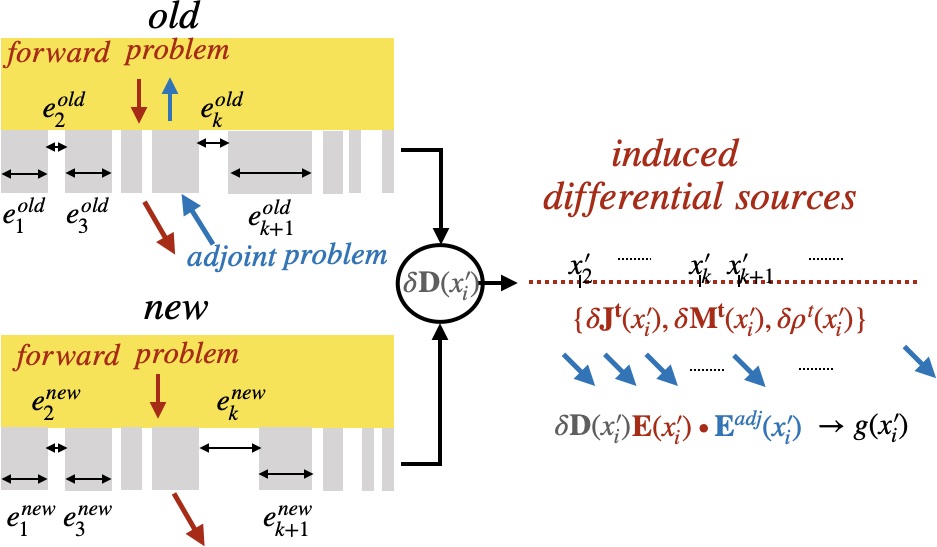}}
  \centering
    \subfigure [\label{TO_lens2}]
        {\includegraphics[width=.7\textwidth]{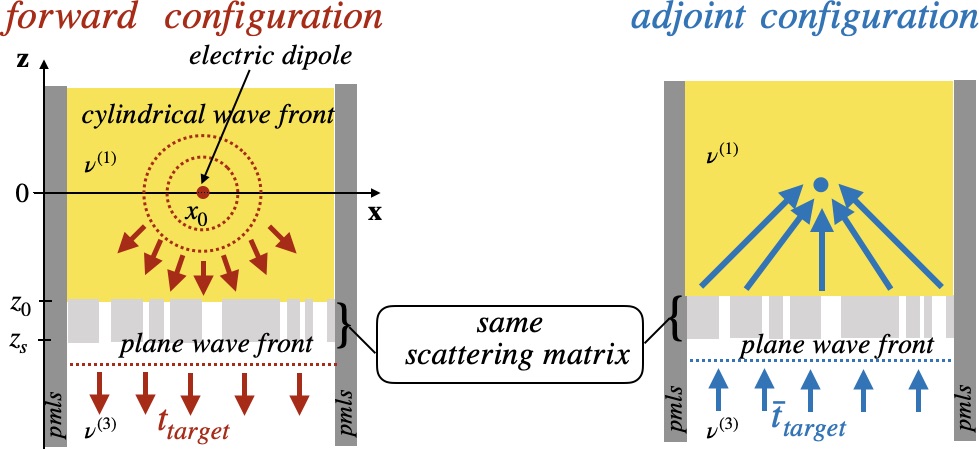}}

\caption{\label{TO_DIRECT_ADJOINT_TO_lens2}
Fig.\ref{TO_DIRECT_ADJOINT}: Sketch  of the direct and the adjoint problems used to compute the gradient of the FOM. Principle of the adjoint based method : the variation of the permittivity function of the old structure induces fictitious sources whose variations are considered to be locally proportional to the variation of the electric density vector. Fig.\ref{TO_lens2}: Design of a one dimensional dielectric metalens: here, $(x_f,z_f)=(x_0,0)$. The adjoint-based topology optimization method involved two different sources: an electric dipole and a polarized incident plane wave. Doing so, no curvilinear phase profile assumption is enforced during the optimization process} 
\end{figure}

Let us denote by $\mathcal{I}$ the design domain on which the discontinuous permittivity function $\varepsilon(x)$ is described by a piecewise constant function. 
The sub-intervals where $\varepsilon (x)$ is constant will be denoted by $\mathcal{I}_k=[x_{k-1},x_k]_{1\leq k\leq N_{p}}$.
In the proposed method, each sub-interval $\mathcal{I}_k$ is associated with a fitness $g_k^{(t)}$ defined from the function $g^{(t)}$ as the normalized first moment of $g^{(t)}$ on $\mathcal{I}_k$ or as a normalized average value of the border-location variable $x$ on the sub-interval $\mathcal{I}_k$  weighted according to the gradient of the figure of merit:
\begin{equation}\label{barycenter}
    g_k^{(t)}=\dfrac{\int_{\mathcal{I}_k^{(t)}} x g^{(t)}(x) dx}{\int_{\mathcal{I}_{k}^{(t)}} x dx} 
\end{equation}
At each iteration, the sequence of fitness values $g_k^{(t)}$ are sorted (descending) and used to search the location  of each variable $x_k$ leading to the FOM improvement. 
\subsection{Best current pattern update}
In this phase, the fitness $g_k^{(t)}$ is used to perturb the value of the variable $x_k$ which can undergo an ascending or descending increment by taking into account the minimum size constraints. Only the increment direction leading to the best result is kept as the new optimal $x_k$ location:
\begin{equation}\label{update_xk}
    x_k^{new}=best\left\{x_k^{(t)}-\delta x_k^{(t)}, x_k^{(t)},x_k^{(t)}+\delta x_k^{(t)}\right\} 
\end{equation}
With $\delta x_k^{(t)}=\alpha^{(t)}g_k^{(t)}$ where  $\alpha^{(t)}=a_0 atanh(1-t/t_{max}) = a_0\beta^{(t)}$ is a parameter decreasing  to zero with respect to the number of iterations, $t_{max}$ denotes the maximal number of iterations. The constant $a_0$ is the rate at the first iteration. Once the best current sequence of variables $x_k^{new}$ are identified, a new sequence of the nano-rods widths is computed through: 
\begin{equation}
 e_k^{new}=x^{new}_{k}-x^{new}_{k-1}, \quad k\in[1,N_p], \quad x_0=0,  \quad x_{N_p}=d   
\end{equation}
At this point, random oscillations are applied to the best current result, namely, the vector $ [e_k^{new}] $, via a random contraction or dilatation  mechanism. This random oscillatory behavior is mathematically formalized as follows:
\begin{equation}\label{stressed_variable}
    e_k^{new}+\beta_0\beta^{(t)}r_k
\end{equation}
where $[r_k]=2rand(N_p,1)-1$ is a vector of random variables in the interval  $[-1,1]^{N_p}$, simulating an uncertainty of the oscillation mode of the current vector $[e_k^{new}]$. A parameter $\beta_0$ is added for a fine-tuning  of the rods and air-gaps widths $e_k^{new}$ so that the induced perturbations do not change so much their values. In this paper $\beta_0$ is  set to $\beta_0=min_k([e^{new}_k])/\eta$. The parameter  $\eta$ is set to $5$ in this paper.  
Finally, Eq. \ref{stressed_variable} is used to update the previous ({\it old}) geometry as follows:
\begin{equation}
e_k^{old}=e_k^{new}+\beta_0\beta^{(t)}r_k    
\end{equation}
and the whole process restarts iteratively, by gradually improving the FOM until it converges to a final structure.
\section{Results and Discussion}
We are now ready to prove the efficiency of the concept proposed in this paper through two numerical examples related to two inverse problems. First we will analyse the performance of the method in the design of a metagrating that deflects a normally-incident TM-polarized plane wave, with wavelength $\lambda$ onto a given transmittance angle $\theta_d$ with the highest intensity. 
Second, the design of an aperiodic metasurface (a metalens) is investigated. Generally, inverse problems are ill-posed in the sense that the solutions are highly sensitive to the initial conditions. Contrary to a well-posed problem, to achieve a stable numerical solution, an ill-posed problem requires to be regularized (i.e. restated by including additional assumptions). In the inverse design problems under consideration, this process may consist in limiting the number of nano-rods of the designed metasurfaces. Hence, in the case of a one-dimensional metasurface, we state the inverse design problem as follows: 
find the best way to achieve a high-performance structure by
fine-tuning a sequence of key parameters; namely the size and the relative positions of a given number of randomly distributed nano-rods with random sizes (enforcing, possibly, minimum size constraints). 
\subsection{Metagrating design }
\begin{figure}[htb!]
 \centering
        {\includegraphics[width=0.5\textwidth]{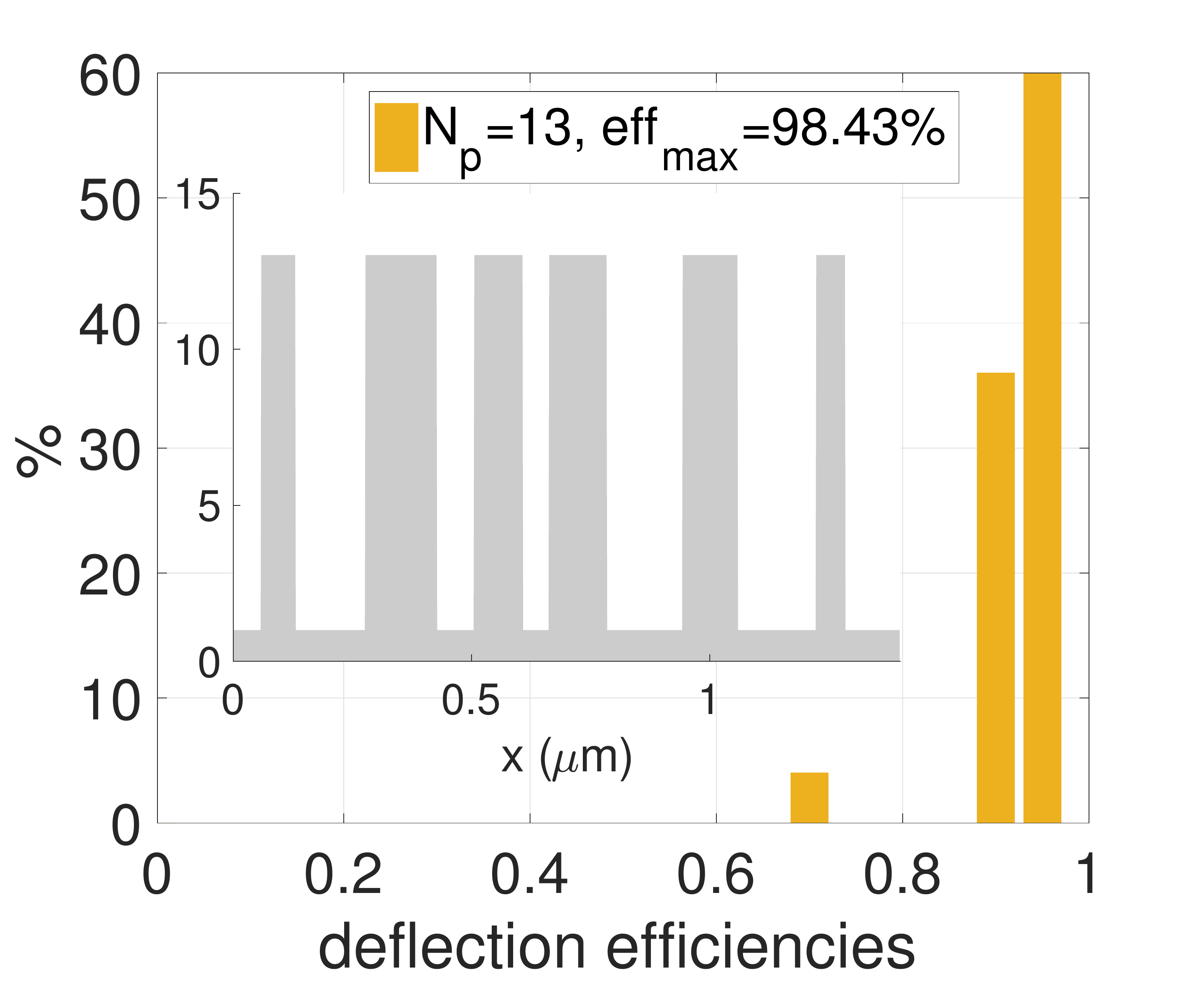}}
\caption{\label{device_lamb09_teta40_Np13}
Histogram of efficiencies and best high transmission device deflecting a TM incident $0.9 \mu m$-wavelength plane wave into $40^o$ transmission angle. The optimized metagrating consists of six nano-rods and seven air-gaps.}
\end{figure}

In the first part of this study, we consider a one-dimensional metagrating, consisting of Si nano-rods with refraction index $3.6082$, deposited on an SiO$_2$ substrate (refractive index : $1.45$). The  grating's height is set to $h=650 nm$. 
To comply with standard fabrication techniques, a minimum size of both rods and air-gaps widths are set to $50 nm$ within the optimization process. Here the FOM will be the efficiency diffracted in the desired direction. 
Two wavelengths, $\lambda=0.9 \mu m$ and $\lambda=1.1 \mu m$; and two deflection angles $\theta_d=40^o$ and $\theta_d=80^o$ are investigated. The parameters $e_{min}$ and $e_{max}$ of Eq. \ref{width1} are set to $50 nm$  and $100 nm$ respectively.  
\begin{figure}[htb!]
 \centering
   \subfigure [\label{compare_histo_lamb09_teta40} histograms of optimized devices]
        {\includegraphics[width=0.35\textwidth]{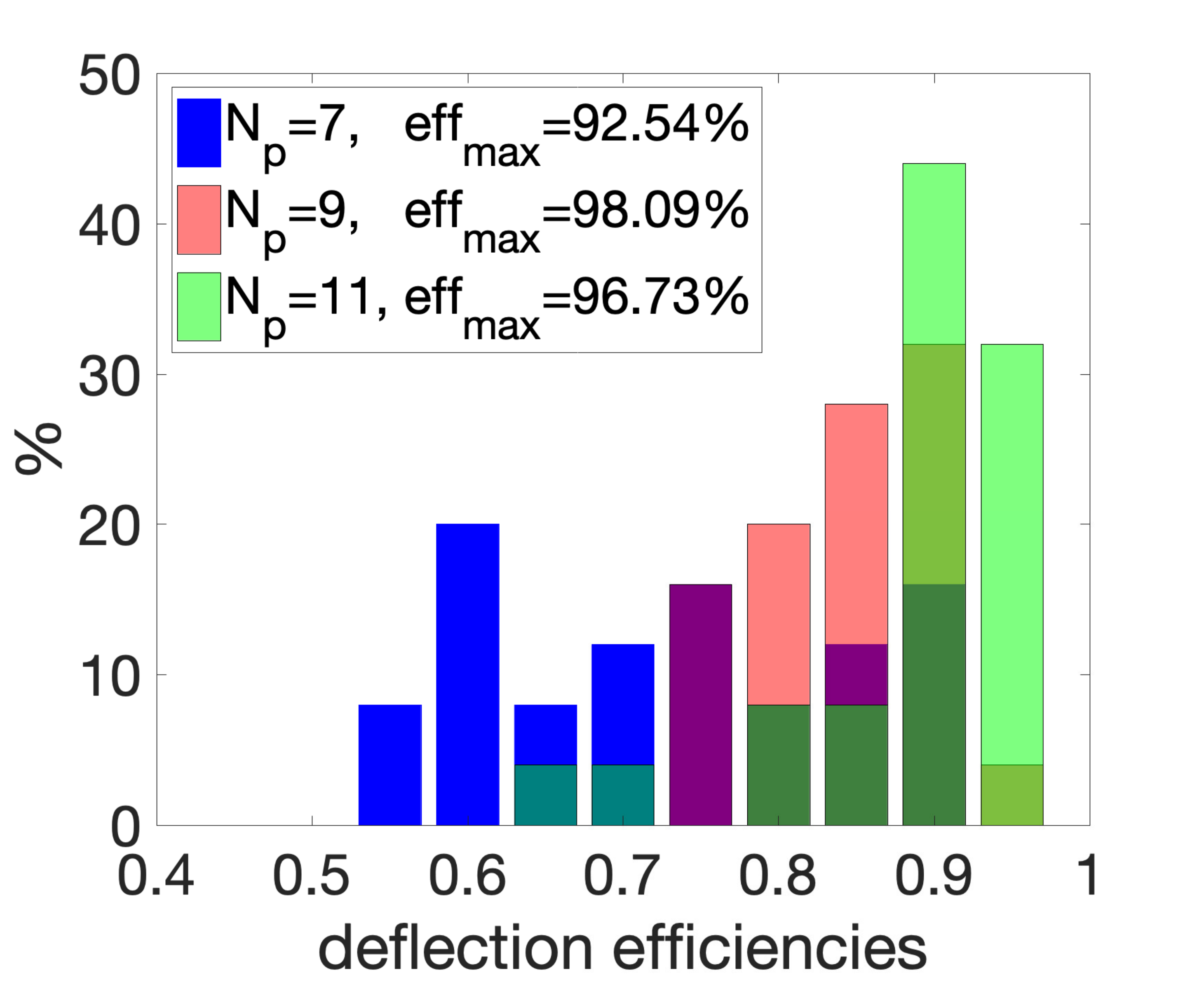}}
 \centering
   \subfigure [\label{device_lamb09_teta40_np7} the best device for $N_p=7$]
        {\includegraphics[width=0.35\textwidth]{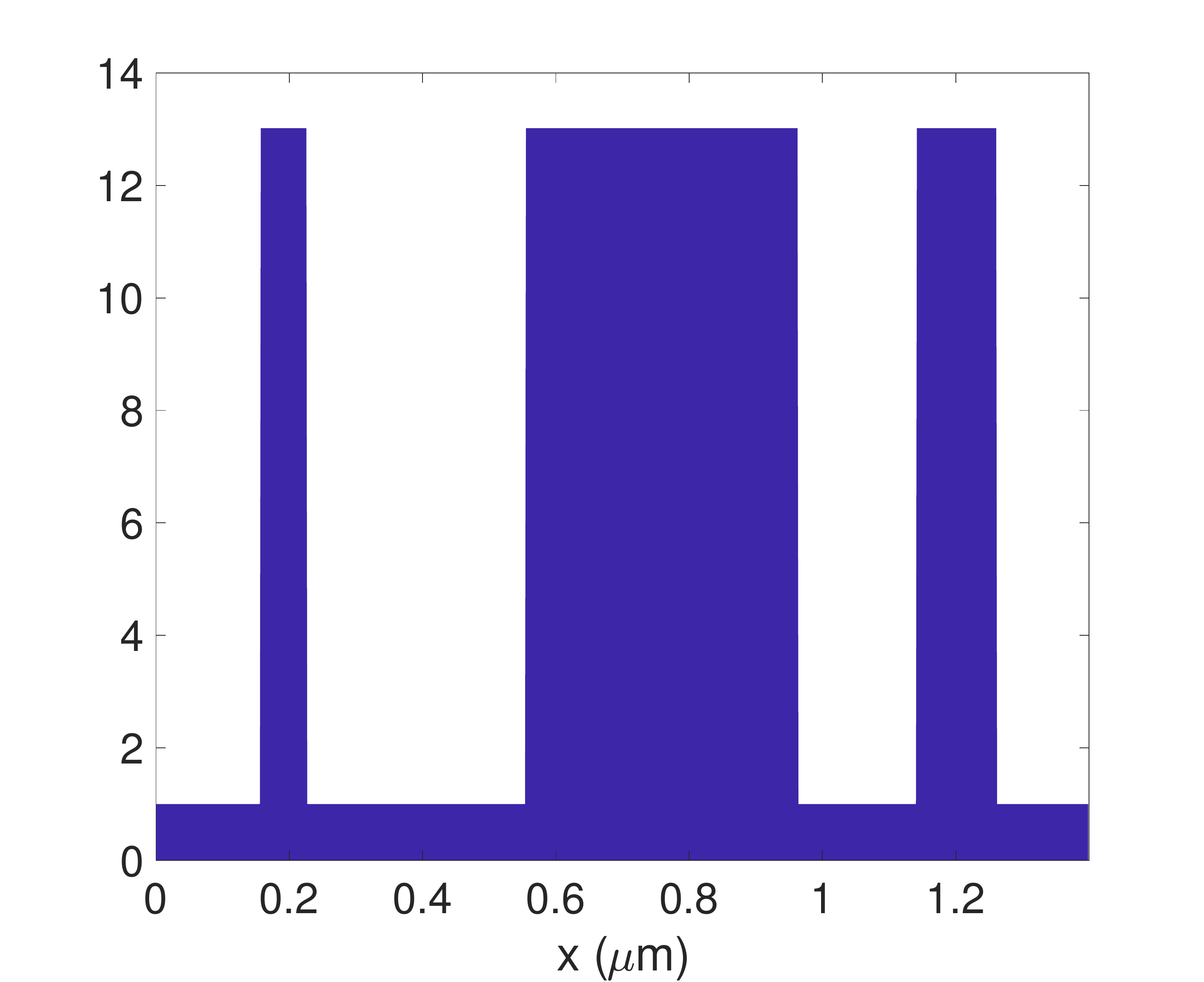}}
        \centering
   \subfigure [\label{device_lamb09_teta40_np9} the best device for $N_p=9$]
        {\includegraphics[width=0.35\textwidth]{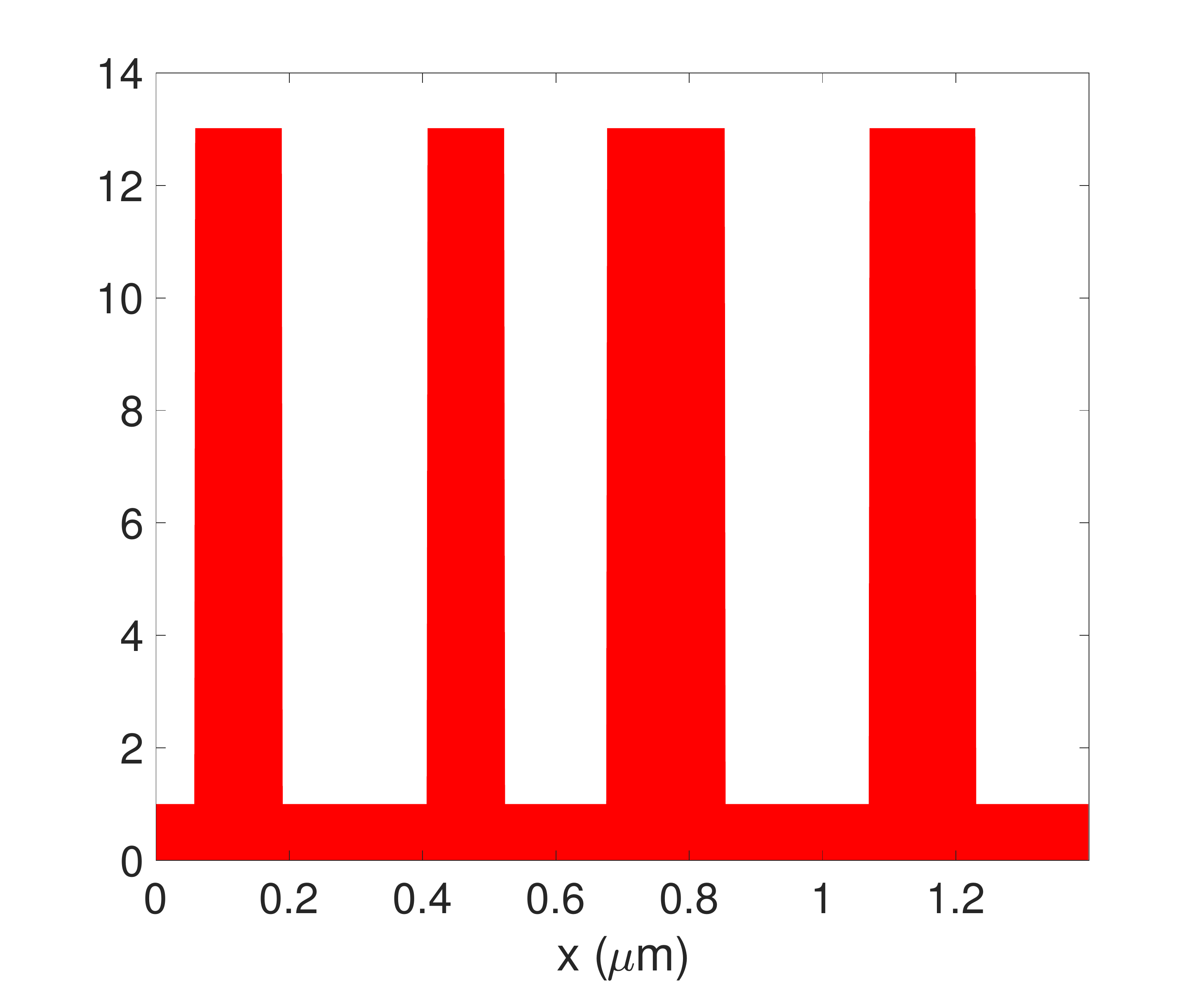}}
 \centering
   \subfigure [\label{device_lamb09_teta40_np11} the best device for $N_p=11$]
        {\includegraphics[width=0.35\textwidth]{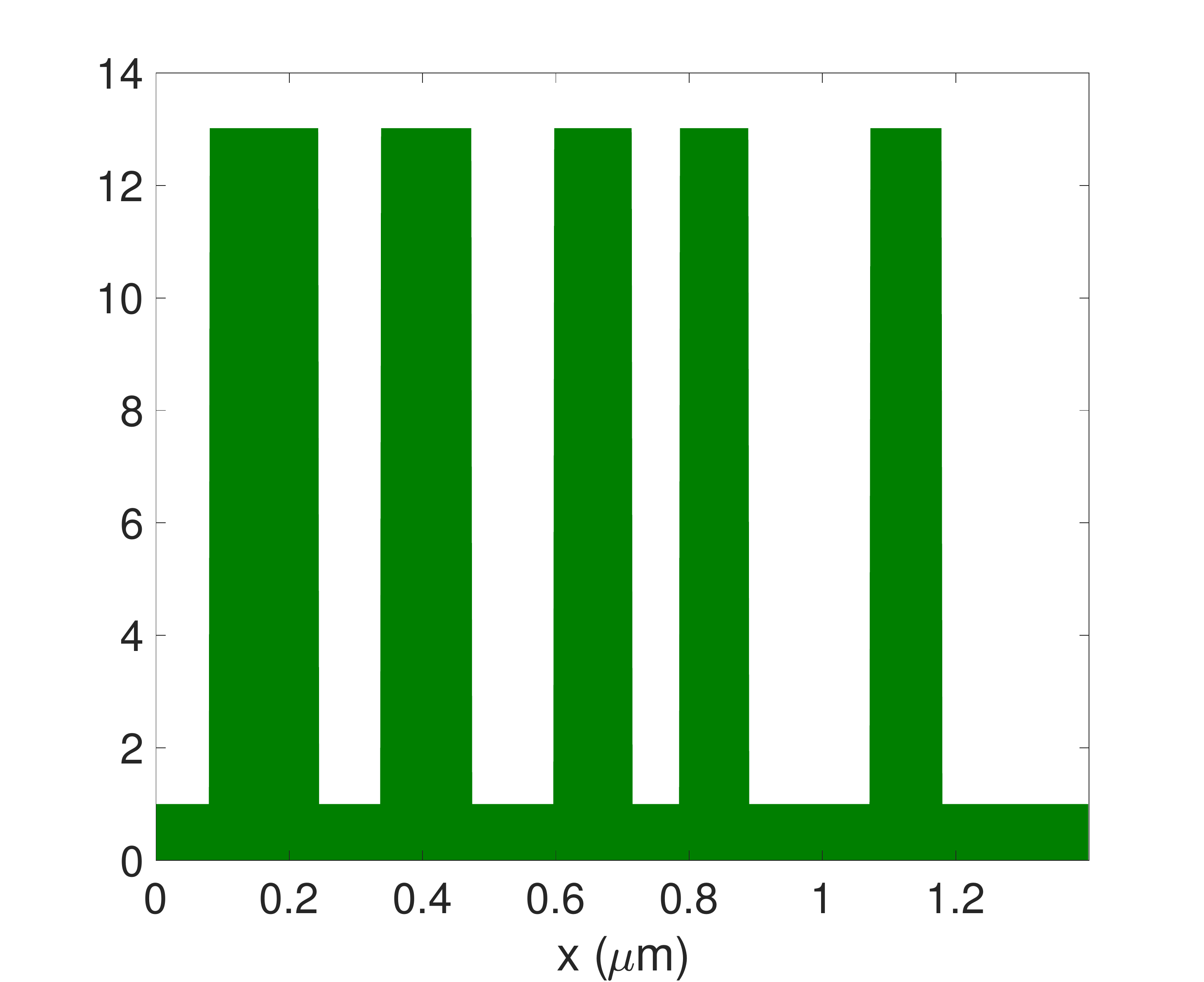}}
\caption{\label{device_lamb09_teta40_np_7_9_11}
Figure \ref{compare_histo_lamb09_teta40}: Efficiencies histograms of optimized metagratings, for three values of the parameter $N_p$: $N_p=7$ ($3$ nano-rods + $4$ air-gaps), $N_p=9$ ($4$ nano-rods + $5$ air-gaps) and $N_p=11$ ($5$ nano-rods + $6$ air-gaps). Figures \ref{device_lamb09_teta40_np7}, \ref{device_lamb09_teta40_np9} and \ref{device_lamb09_teta40_np11}: sketch of  best devices obtained from $3$, $4$ and $5$ nano-rods respectively. Numerical parameters: $\lambda=0.9\mu m$, deflection angle of $40^o$, polarization TM.}
\end{figure}
Figure \ref{compare_histo_lamb09_teta40} presents the histograms showing the distribution (according to  diffraction efficiency) of optimized metagratings deflecting a normally incident plane wave at  $\lambda=0.9\mu m$ into the direction of $40^o$, for three values of the parameter $N_p$: $N_p=7$ ($3$ nano-rods + $4$ air-gaps), $N_p=9$ ($4$ nano-rods + $5$ air-gaps) and $N_p=11$ ($5$ nano-rods + $6$ air-gaps).
Twenty five sequences of $N_p$-tuple random  variables $([e_k^{old}],[e_k^{new}]) \in [50nm,100nm]^{N_p}\times [50nm,100nm]^{N_p}$ are initially  generated, and for each couple of initial profiles, $100$ iterations are performed in the optimization process. As expected, for the chosen couple $(\lambda, \theta_d)=(0.9\mu m,40^o)$, high performance solutions are reached whatever the $N_p$ values.
The efficiencies histograms of Figure \ref{compare_histo_lamb09_teta40} 
show that this latter systematically narrows when $N_p$ increases, indicating that the search of the optimal device performs better when $N_p$ increases. 
In other words, the part of initial candidates yielding high transmission devices increases with the number of nano-rods. 
However, the best 4NR device (i.e. device with four nano-rods) is slightly better than that of the best 5NR device. This indicates that the probability to obtain the most efficient device does not systematically and monotonously increase with $N_p$ even though the probability to reach the basin of favorable solutions increases with $N_p$.
Figure \ref{device_lamb09_teta40_Np13}
presents the histogram of efficiencies of optimized metagratings with six nano-rods and seven air-gaps ($N_p=13$). As expected, the efficiency histogram is narrower than those of the previous cases since $N_p$ is higher. This device exhibits a slightly higher transmission efficiency eff$_{max}=98.43\%$ than the 4NR device which yields eff$_{max}=98.09\%$. \\
\begin{figure}[htb!]
 \centering
   \subfigure [\label{device_lamb11_teta80_npZ} the best device for $N_p=7$]
        {\includegraphics[width=0.35\textwidth]{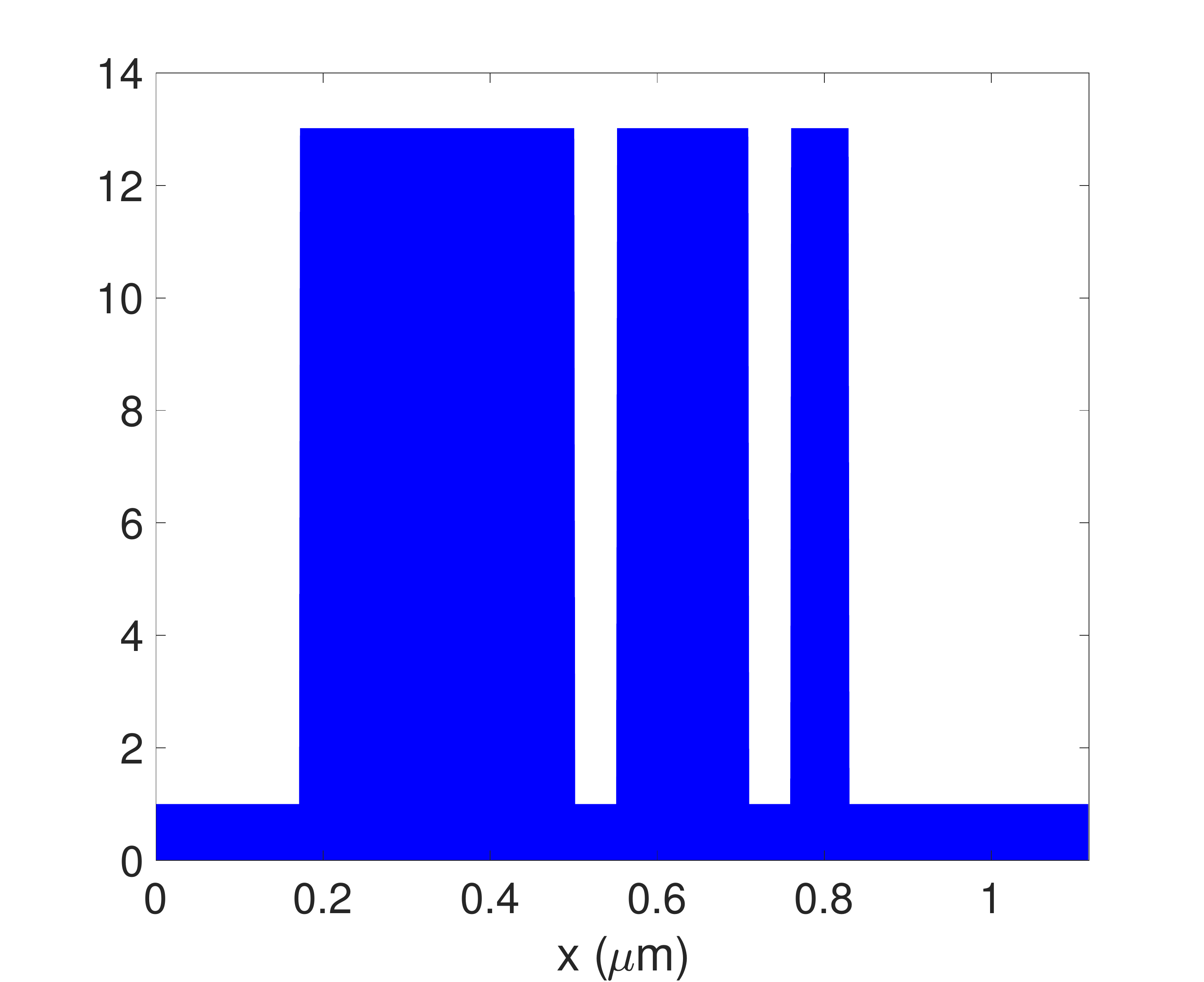}}
        \centering
   \subfigure [\label{device_lamb11_teta80_np9} the best device for $N_p=9$]
        {\includegraphics[width=0.35\textwidth]{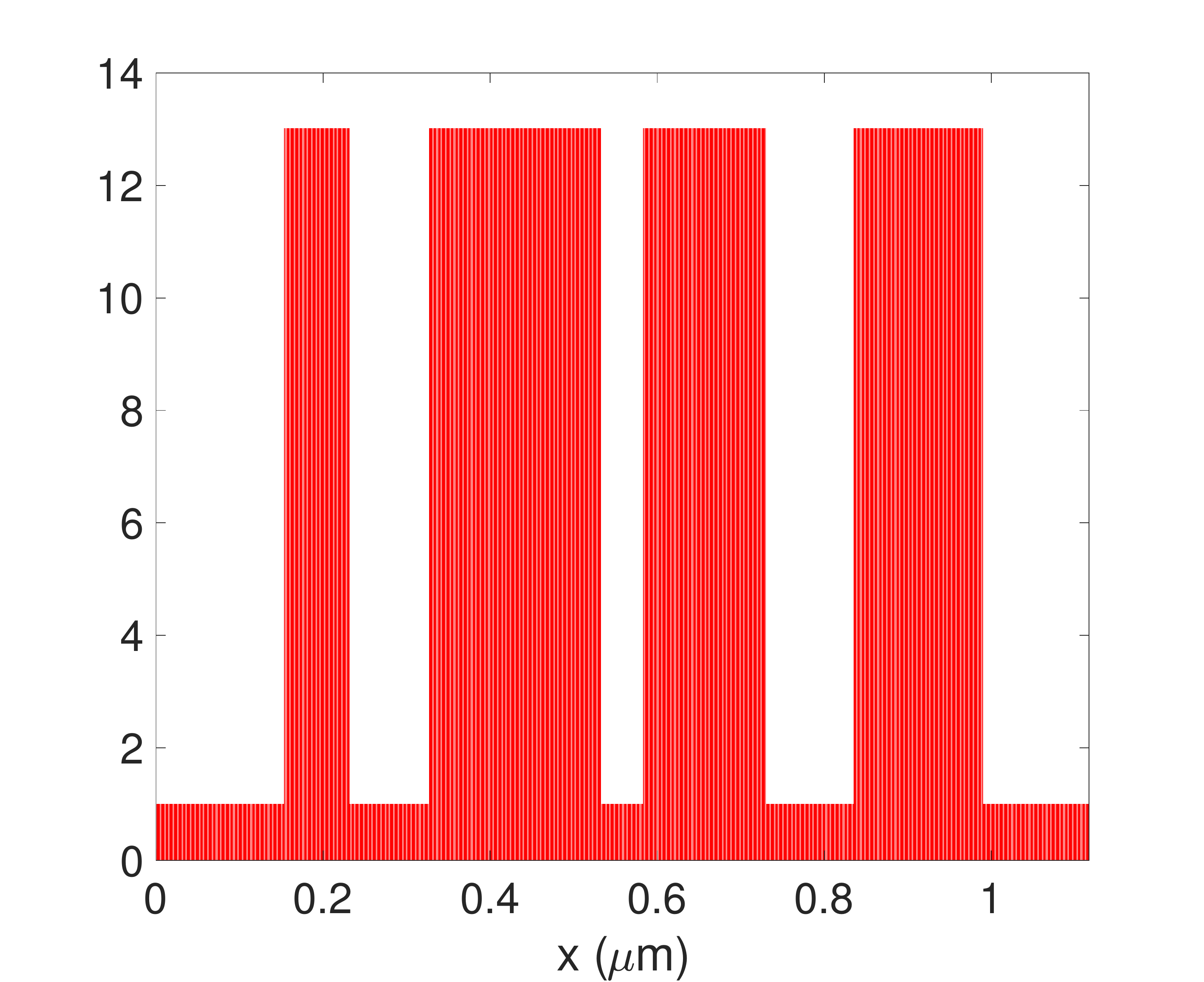}}
 \centering
   \subfigure [\label{device_lamb11_teta80_np11} the best device for $N_p=11$]
        {\includegraphics[width=0.35\textwidth]{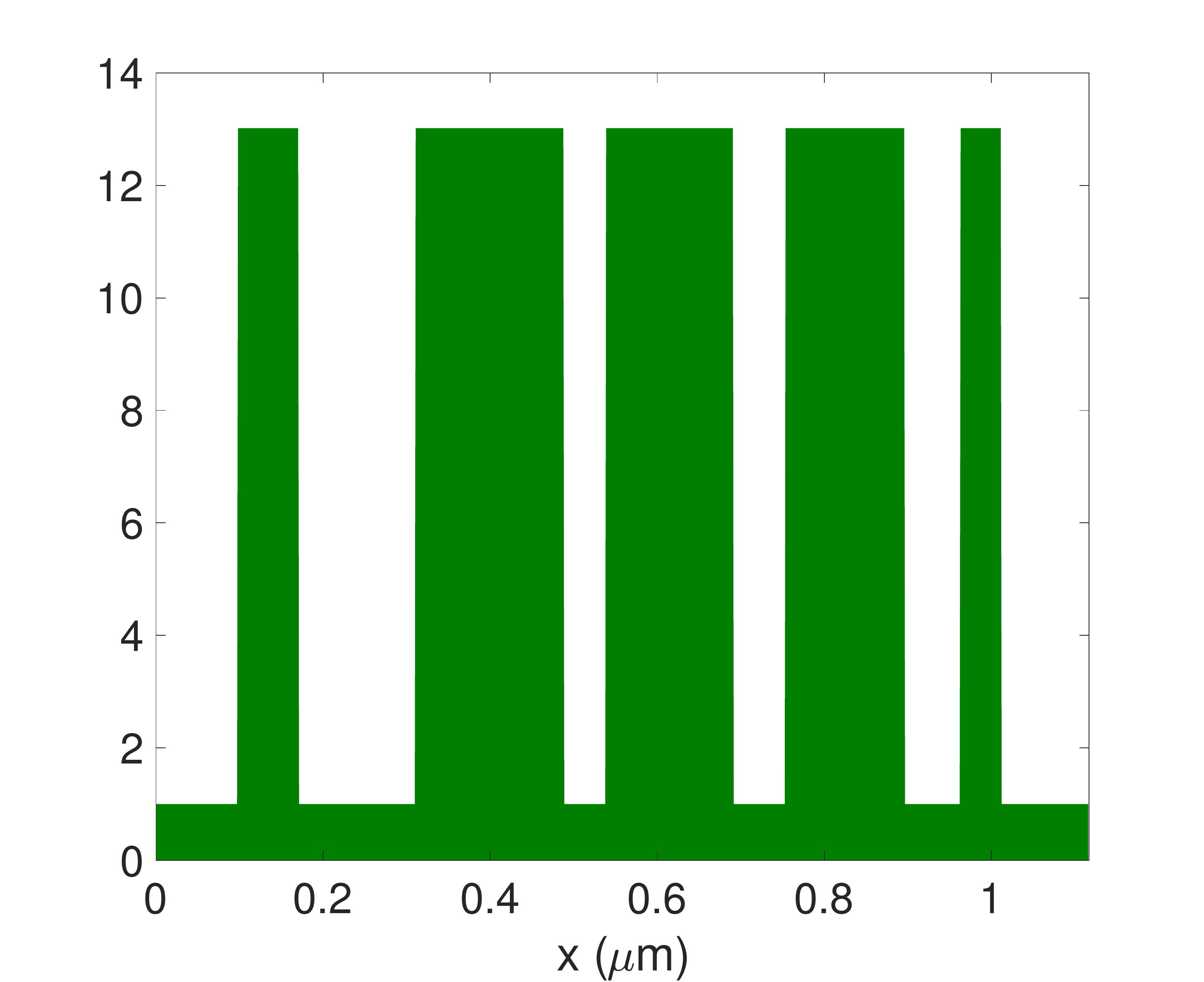}}
        \centering
   \subfigure [\label{compare_histo_lamb11_teta80} histograms of optimized devices]
        {\includegraphics[width=0.35\textwidth]{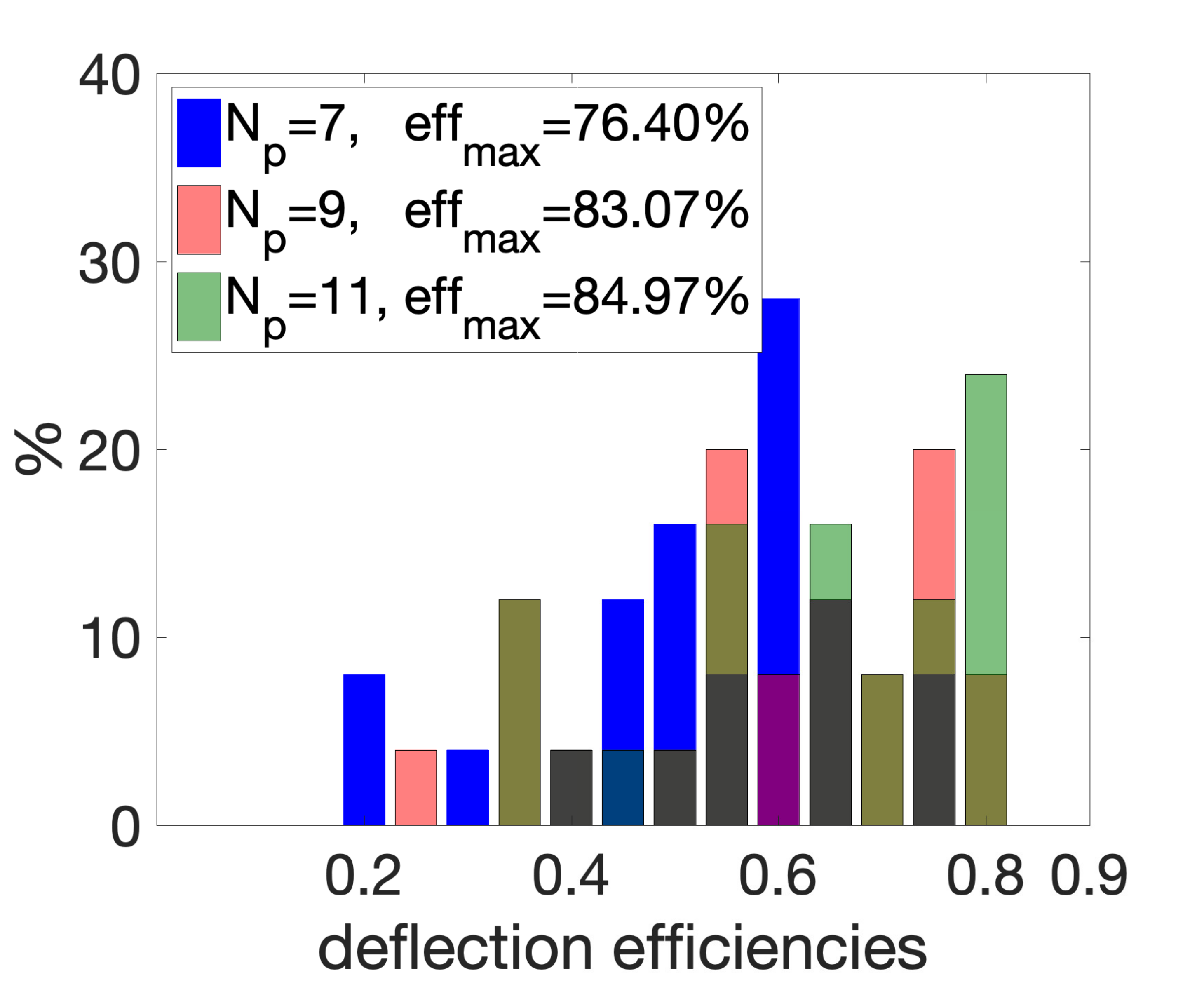}}
\caption{\label{device_lamb11_teta40_np_7_9_11}
 Figures \ref{device_lamb09_teta40_np7}, \ref{device_lamb09_teta40_np9} and \ref{device_lamb09_teta40_np11}: Sketch of best devices obtained for three values of the parameter $N_p$: $N_p=7$ ($3$ nano-rods + $4$ air-gaps), $N_p=9$ ($4$ nano-rods + $5$ air-gaps) and $N_p=11$ ($5$ nano-rods + $5$ air-gaps) respectively. Numerical parameters: $\lambda=1.1\mu m$, deflection angle of $80^o$, polarization TM. Figure \ref{compare_histo_lamb11_teta80}: histograms of optimized metagratings.}
\end{figure}
A complementary analysis of a device operating with a larger deflection angle $\theta_d=80^o$ (which is considered to be a harder design to achieve) and at $\lambda=1.1 \mu m$ is presented in Figure \ref{device_lamb11_teta40_np_7_9_11}. Three kinds of structures are simulated: a 3NR device whose best result is presented in Figure \ref{device_lamb11_teta80_npZ}, a 4NR device displayed in Figure \ref{device_lamb11_teta80_np9} and a 5NR device showed in Figure \ref{device_lamb11_teta80_np11}. 
The comparison between the efficiencies histograms of these three kinds of structures is presented in Figure \ref{compare_histo_lamb11_teta80}. 
Here again, the histograms narrow when $N_p$ increases. In the current example, we can remark that, with an efficiency close to $85\%$, the  5NR device seems to be the most efficient one operating with the couple $(\lambda, \theta_d)=(1.1\mu m,80^{o})$.\\
The computation times for the design of metagratings  deflector using:
\begin{itemize}
    \item the Fourier modal method (FMM)\cite{Knop, Granet_Guizal,Lalanne, Li, Granet_ASR} (in its conical formulation), implemented with $2\times 40 +1$ Fourier basis functions (the size of the eigenvalues equations matrix is then $2(2\times 40 +1)\times2(2\times 40 +1)$ )
    \item 25 randoms initial conditions
    \item 100 iterations for each initial conditions
\end{itemize} 
are about $3.9965\times10^3 s$ for $N_p=7$ ($3$ nano-rods), $4.8615\times10^3 s$ for $N_p=9$ ($4$ nano-rods), $5.8217\times10^3 s$ for $N_p=11$ ($5$ nano-rods) and $6.6213\times10^3 s$ for $N_p=13$.  These computations reported on tables \ref{panel_deflector_times}
are performed on a classical Laptop DELL PRECISION 7720, with a processor intel CORE i7 (3.10Ghz).  A CPU time of $6.7\times10^4 s$ is required in the  classical TO using a continuous initial profile function described with $2^8$ pixels. Regarding these times performances, the proposed method is $10$ times less expensive in terms of computation time than the classical TO method based on the use of initial highly pixelated-continuous profile. 
\begin{table}[h!]
\centering
\begin{tabular}{|c||c c c c|}
\hline 
{$N_p$} & 7 & 9 & 11 & 13\\
\hline
\hline
computation times (s)  &  $4\times10^3$ &   $4.9\times10^3$  &   $5.8\times10^3$ & $6.6\times10^3$  \\
\hline
\end{tabular}
\caption{\label{panel_deflector_times}  Panel of the CPU computation times for the design of metagratings  deflectors   using the Fourier modal method (FMM) in its conical formulation, implemented with $2\times40+1$ Fourier basis functions (the size of the eigenvalues equations matrix is then $2(2\times 40 +1)\times2(2\times 40 +1)$, 25 randoms initial conditions, 100 iterations for each initial conditions. 
}
\end{table} 
\subsection{Metalens design}
\begin{figure}[hbt!]
        \centering
   \subfigure [\label{compare_mod_Ey_trans_lamb064_nb52} $|E_y(x,0)|^2/\int_{\mathcal{I}}| E_y^{inc}(x,z_s)|^2 dx$]
        {\includegraphics[width=0.4\textwidth]{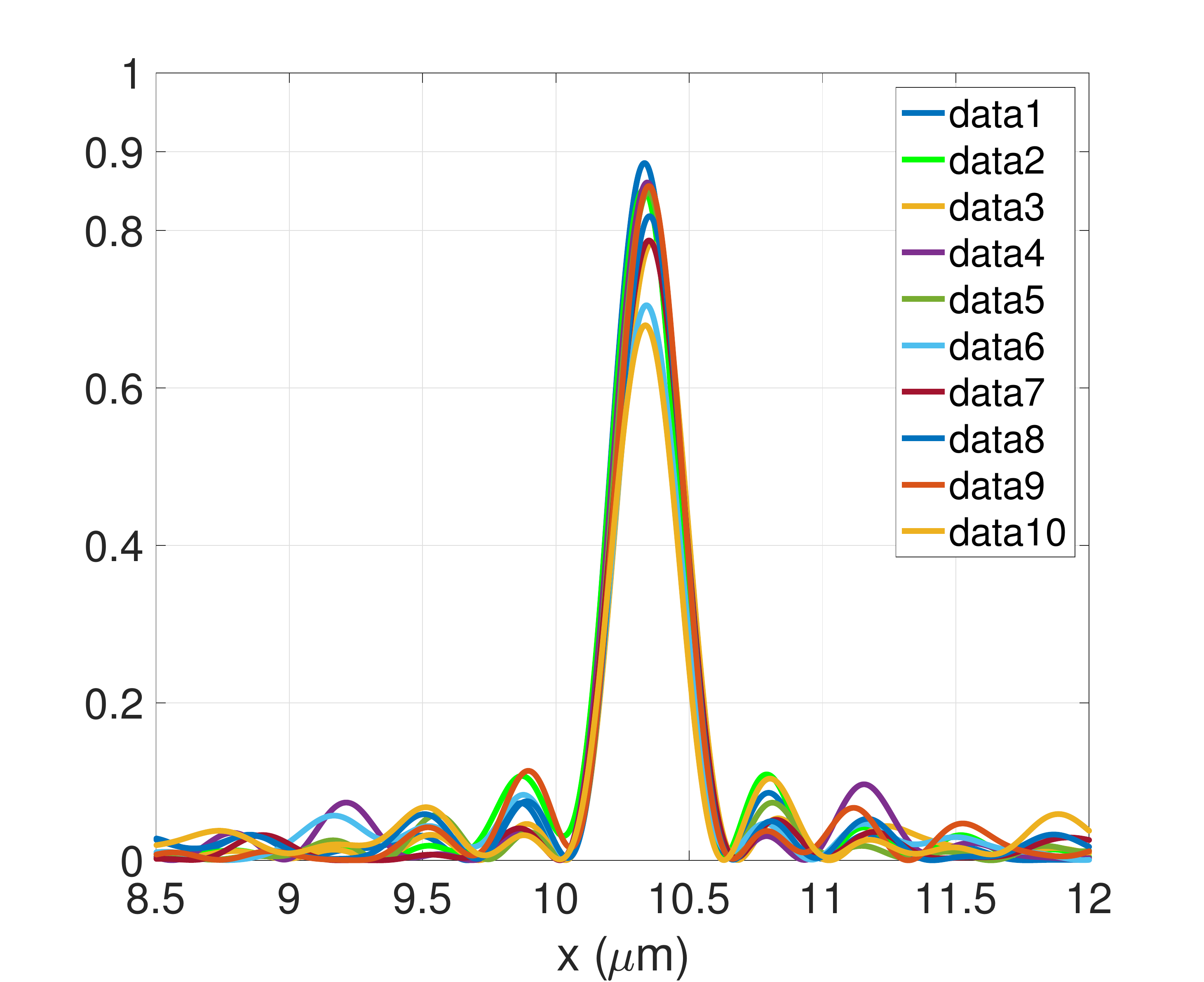}}
 \centering
   \subfigure [\label{compare_mod_Ey_long_lamb064_nb52} $|E_y(x_0,z)|^2/\int_{\mathcal{I}}| E_y^{inc}(x,z_s)|^2 dx$]
        {\includegraphics[width=0.4\textwidth]{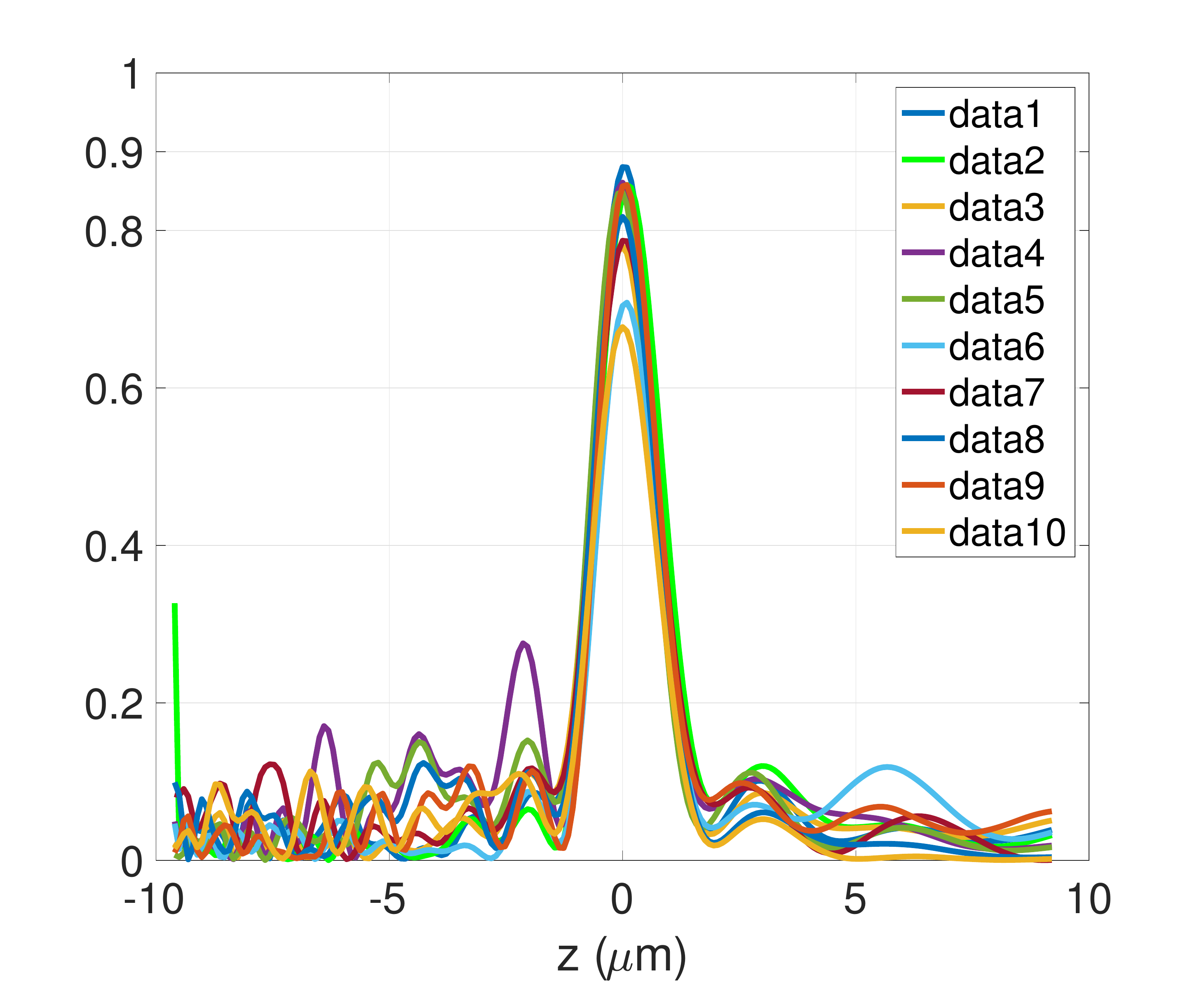}}
\caption{\label{compare_mod_Ey_trans_long_lamb064_nb52}
Design of a $32\lambda$ wide metalens consisting of $N_p=52$ nano-rods and air-gaps focusing a normal incident TE-polarized plane wave with a wavelength $\lambda=0.64\mu m$ at a distance of $15\lambda$.
Line scans of the  normalized  electric field intensity  in the transverse  ($z_f=0$) (Figure \ref{compare_mod_Ey_trans_lamb064_nb52})  and in the longitudinal $x=d/2$ (Figure \ref{compare_mod_Ey_long_lamb064_nb52})  focal plane respectively,  for ten random realizations of initial conditions.   
Numerical parameters: $\lambda=0.64\mu m$, $N_p=52$, height $e_z=325 nm$, TE polarization.}
\end{figure}

\begin{figure}[htb!]
        \centering
   \subfigure [\label{mode_Ey_lambd064_nb52} $|E_y(x,z)|^2/\int_{\mathcal{I}}| E_y^{inc}(x,z_s)|^2 dx$]
        {\includegraphics[width=0.4\textwidth]{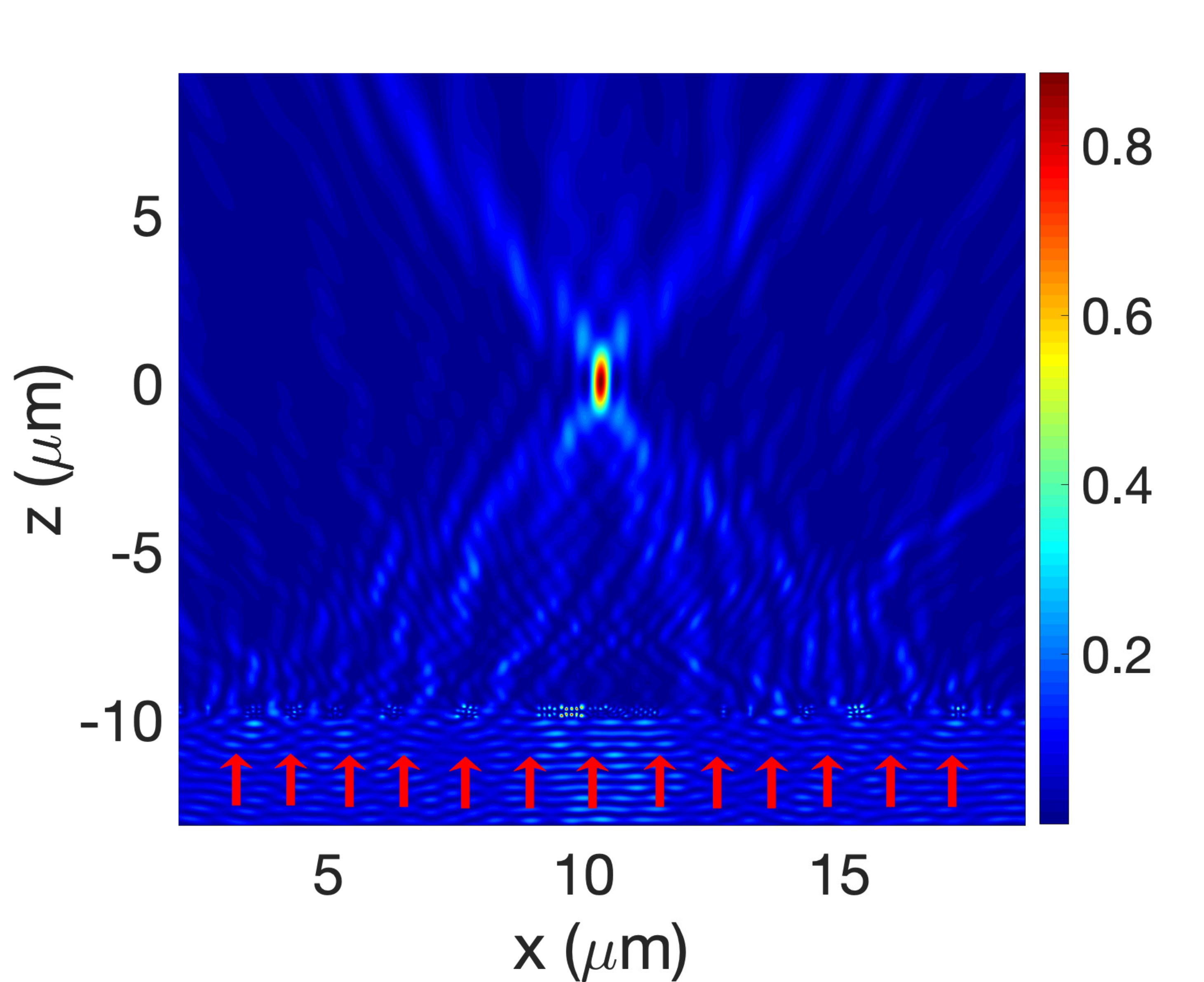}}
 \centering
   \subfigure [\label{angle_Ey_lambd064_nb52} $angle(E_y(x,z))$]
        {\includegraphics[width=0.4\textwidth]{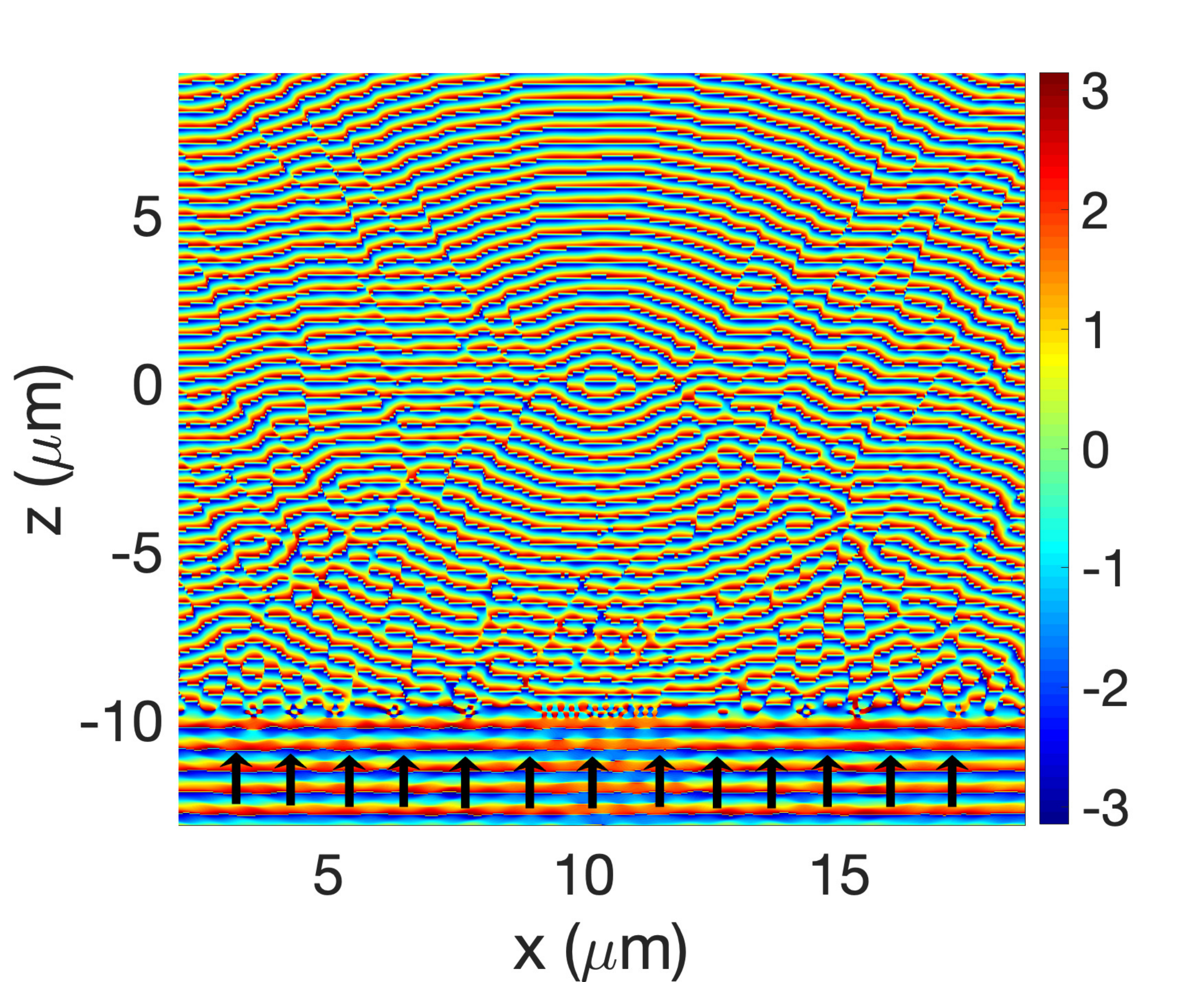}}
\centering
   \subfigure [\label{mod_Ey_trans_lamb064_nb52} $|E_y(x,0)|^2/\int_{\mathcal{I}}| E_y^{inc}(x,z_s)|^2 dx$]
        {\includegraphics[width=0.35\textwidth]{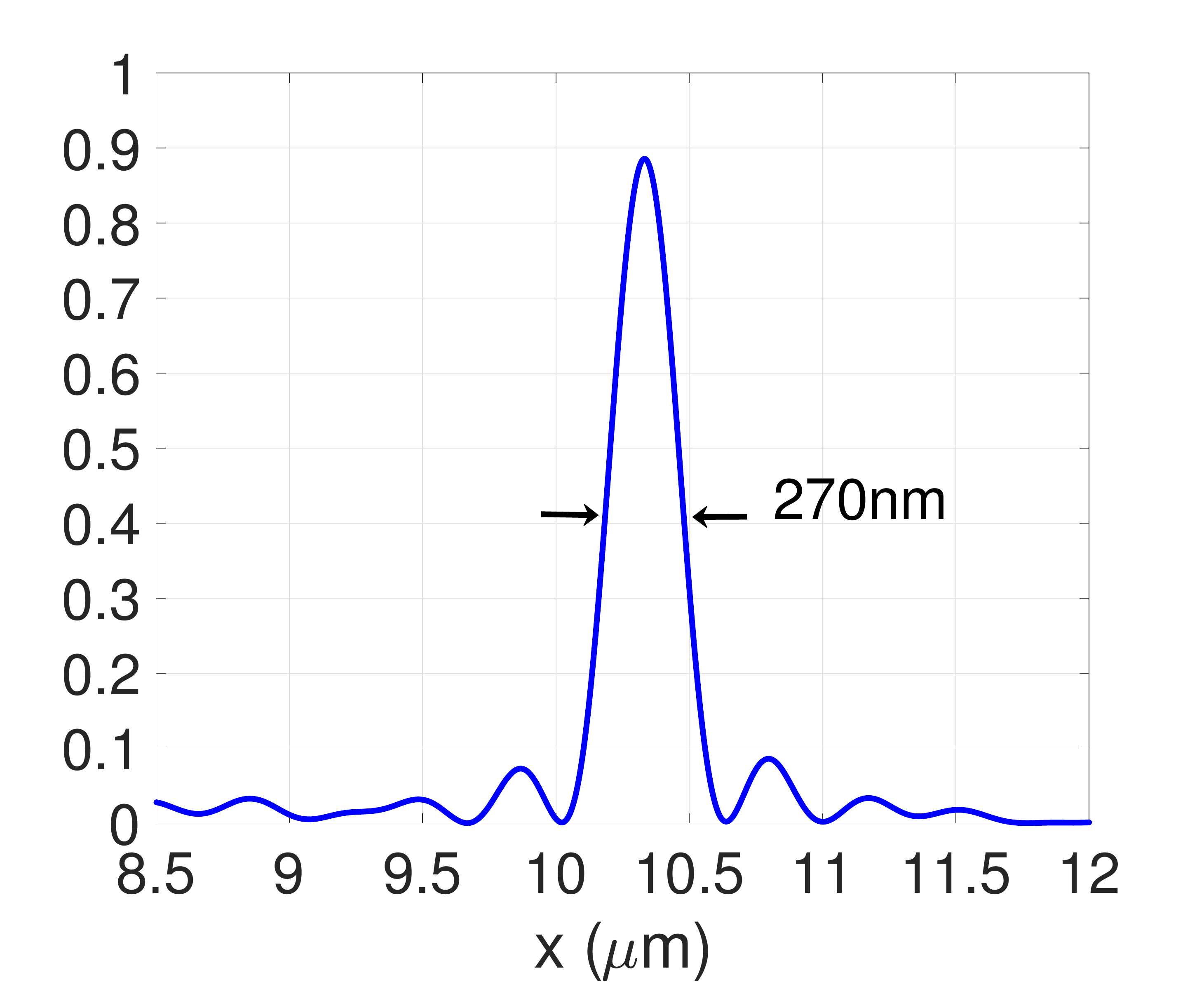}}
        \hspace{1cm}
 \centering
   \subfigure [\label{mod_Ey_long_lamb064_nb52} $|E_y(x_0,z)|^2/\int_{\mathcal{I}}| E_y^{inc}(x,z_s)|^2 dx$]
        {\includegraphics[width=0.35\textwidth]{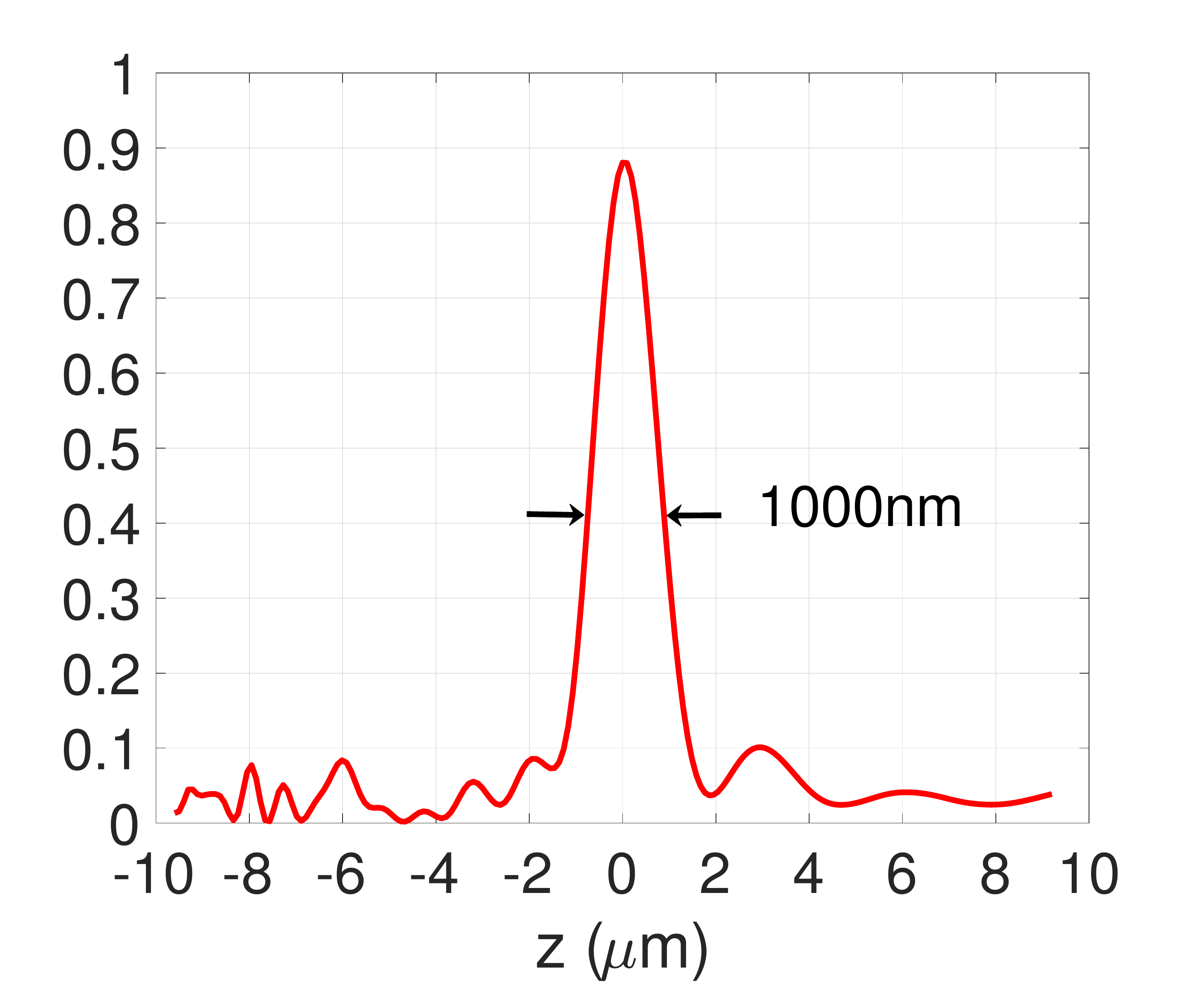}}
 \caption{\label{Ey_long_lamb064_nb52} Design of a $32\lambda$ wide metalens consisting of $N_p=52$ nano-rods and air-gaps focusing a normally incident TE-polarized plane wave, with a wavelength $\lambda=0.64\mu m$, at a distance of $15\lambda$: best optimized structure  results. Figure \ref{mode_Ey_lambd064_nb52}: Cartography  of the  normalized  electric field intensity in the plane $(xoz)$. Figure \ref{mode_Ey_lambd064_nb52}: Cartography  of the  electric field phase in the plane $(xoz)$.   Figure \ref{mod_Ey_trans_lamb064_nb52}: line scan of the  normalized  electric field intensity $(Ox)$-axis at  $z_f=0$. Figure \ref{mod_Ey_long_lamb064_nb52}: line scan of the  normalized  electric field intensity $(Oz)$-axis at  $x=x_0$.   
Numerical parameters: $\lambda=0.64\mu m$, $N_p=52$, height $e_z=325 nm$, TE polarization.}
\end{figure}

As stated above, the reduction in the adjoint-variables number yields a drastic reduction in the computation time, and, thus, enables the optimization of large-scale aperiodic devices in a relative record times, using a classical personal computer. This will be demonstrated in this second example where we consider a one dimensional $325 nm$-height dielectric metalens consisting of a given number of Si nano-rods with refraction index $3.6082$, deposited on SiO$_2$  (refractive index : $\nu^{(3)}=1.45$). The operating wavelength is set to $0.64\mu m$ and two different cases are investigated: a $51\lambda$-wide metalens with $32$ rods/air-gaps and a $52\lambda$-wide metalens with $203$ rods/air-gaps.\\
In the topology-optimization of metalenses, it is common use to introduce a curvilinear phase profile $\phi$ and the symmetry properties of that phase function explicitly reflect on the geometry of the optimized structure. Doing so, the design domain is implicitly restricted to solutions possessing certain properties of symmetry. 
In these optimization methods relying on phase conditions enforcement, \blue{\cite{Lalanne_lens,Yaoyao, Talukdar}}, each voxel of the design area, located around a node $x$ of the metasurface (where $x$ is the distance between the voxel and the center of the lens) is designed to produce a local parabolic phase $\phi(x)=2\pi/\lambda\left[f-\sqrt{f^2+x^2}\right]$ which generally depends on the focal distance $f$.
Since  $\phi(x)$ is symmetric, the optimized final result will be symmetric. Contrary to these phase-enforcement methods, in our approach we optimize the full device in one time and without enforcing any phase conditions. We will demonstrate that the proposed method systematically provides original unexpected asymmetric on-axis metalenses even though under normal illumination.
Rather than stating the optimization problem as the design of metasurface elements  having a desired phase profile and a high level of energy flow focusing,
our design consists in departing from a structure made of a given number of rods placed inside a given space and then finding an optimal layout that focuses and maximizes the energy flow at a given focal point $(x_f,z_f)$ in the Cartesian plane $xoz$.
In terms of adjoint-based topology, our design objective involves the solutions of two electromagnetic problems: a direct (or forward) problem with an electric dipole source (practical details for the numerical treatment of this problem can be found in \cite{plumey}) and an adjoint reciprocal problem with an incident plane wave (cf. Figure \ref{TO_lens2}). 
We begin with the design of a metalens capable of focusing a normally incident TE plane wave (with $\lambda=0.64\mu m$), traveling the the vacuum ($\nu^{(1)}=1$) at a focal distance $f=15\lambda$ located in the $SiO_2$ medium ($\nu^{(3)}=1.45$). For that, we consider a structure of width $d=32\lambda$, comprising $N_p=52$ nano-rods and air-gaps and perform ten optimizations with various random initial sequences of rods and air-gap widths $[(e^{old}_k,e^{new}_k)]_k$, generated according to Eqs. \ref{width1} and \ref{width2}. For each initial condition, the maximum number of iterations $t_{max}$ is set to $50$.  The minimum size constraint of both rods and air-gaps widths is set to $60 nm$. The Aperiodic Fourier Modal Method (AFMM) is used to treat the forward and the adjoint problems in one simulation thanks to the fact that both of them involve the same scattering matrix (see figure \ref{TO_lens2}). To ensure the convergence of the Fourier series, the permittivity and the electromagnetic field are approximated with $2M+1$ Fourier basis functions, $M=200$ being the so-called truncation order.    
The numerical results are presented  in Figure \ref{compare_mod_Ey_long_lamb064_nb52} where we can see the line scans of the normalized electric field intensity along the $x$-axis at $z_f=0$ (Figure \ref{compare_mod_Ey_trans_lamb064_nb52}) and along the $z$-axis at $x=x_0=d/2$ (Figure \ref{compare_mod_Ey_long_lamb064_nb52}) respectively and for ten different random initial conditions. Although the adjoint based algorithm is coupled with a gradient-based method, which is known to be very sensitive to the initial conditions, a very large part of the ten designed metalenses maintain satisfactory high efficiencies : i.e. exceeding $78\%$.
Now we select the best result and plot, on Figure \ref{mode_Ey_lambd064_nb52}, the normalized electric field intensity $|E_y(x,z)|^2/\int_{I}| E_y^{inc}(x,z_s)|^2 dx$ in the $(X,Z)$-plane, with the corresponding phase distribution in Figure \ref{angle_Ey_lambd064_nb52}. As Figure \ref{mode_Ey_lambd064_nb52} shows, a well-defined and strong focus spot in both the $x$ and $z$ directions, with very weak side-lobes, can be clearly distinguished. The wave-front conversion between the incident wave (with plane wave-front), and a cylindrical wave-front is clearly demonstrated on Figure \ref{angle_Ey_lambd064_nb52}. Figures \ref{mod_Ey_trans_lamb064_nb52} and \ref{mod_Ey_long_lamb064_nb52}, give more insight on the side lobes and on the size of the focal spot. The intensity ratio side-lobe/central-lobe is around $10.5\%$ in the $x$ direction. The LFWHM (Lateral Full Width Half Maximum) is equal to $270nm \simeq 0.42\lambda$ (Fig. \ref{mod_Ey_trans_lamb064_nb52}) while in the axial direction the AFWHM (Axial Full Width Half Maximum) is equal to $1000nm \simeq 1.56\lambda$ yielding a high-quality cylindrical focal spot.
\\

\begin{figure}[htb!]
        \centering
   \subfigure [\label{mode_Ey_lambd064_nb103_Px52lambda} $|E_y(x,z)|^2/\int_{\mathcal{I}}| E_y^{inc}(x,z_s)|^2 dx$]
        {\includegraphics[width=0.4\textwidth]{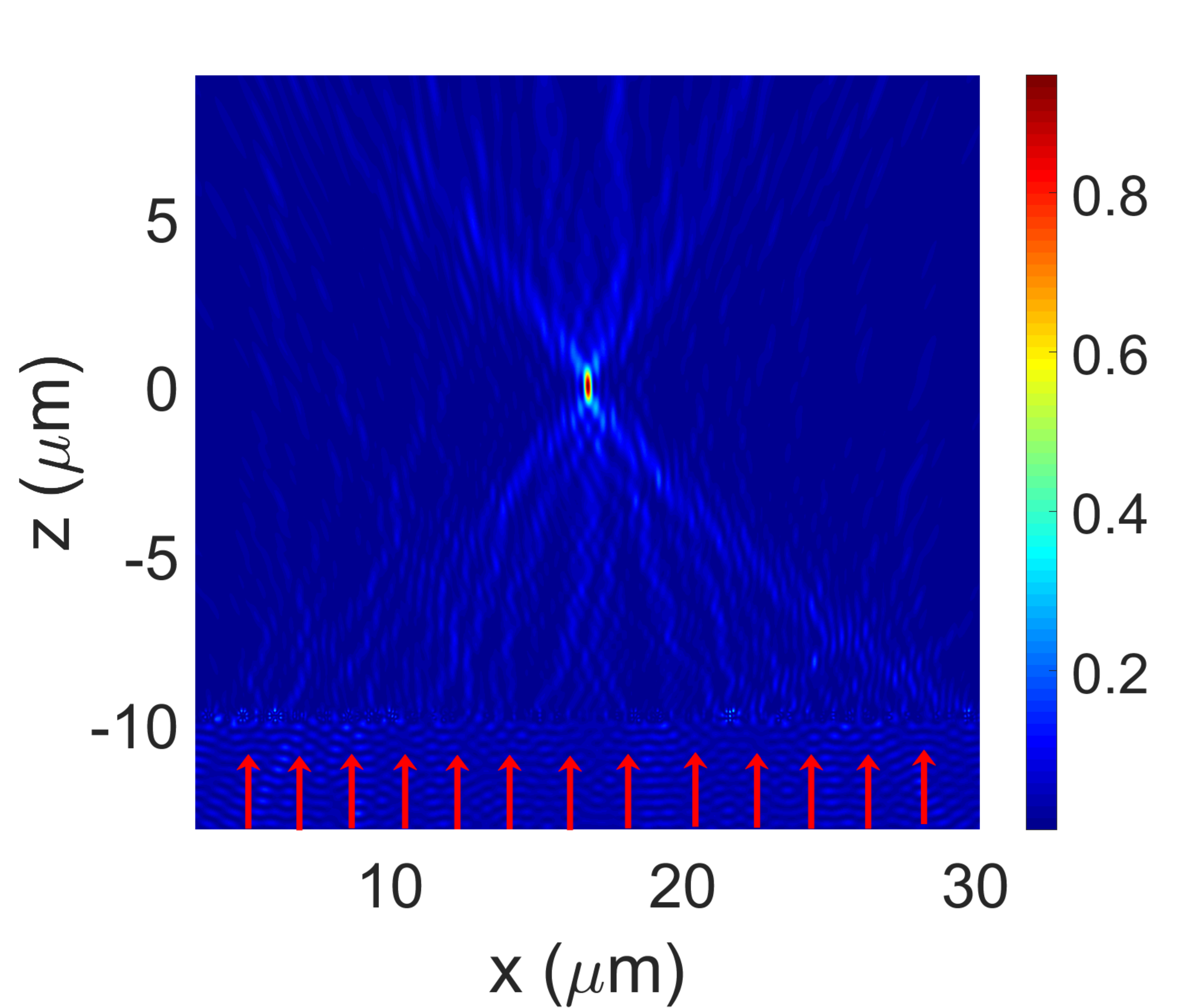}}
 \centering
   \subfigure [\label{angle_Ey_lambd064_nb103_Px52lambda} $angle(E_y(x,z))$]
        {\includegraphics[width=0.4\textwidth]{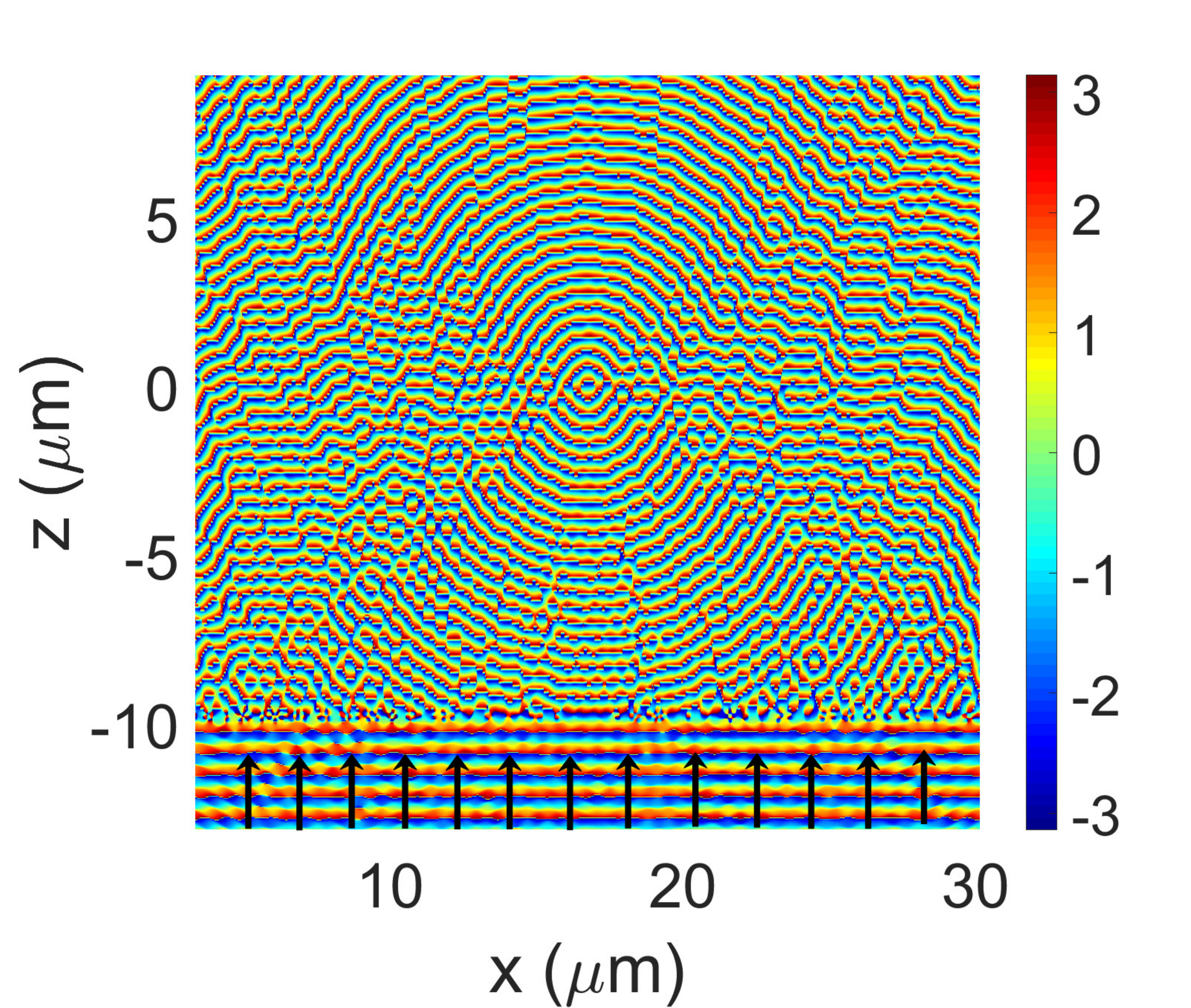}}
\centering
   \subfigure [\label{mod_Ey_trans_lamb064_nb103} $|E_y(x,0)|^2/\int_{\mathcal{I}}| E_y^{inc}(x,z_s)|^2 dx$]
        {\includegraphics[width=0.35\textwidth]{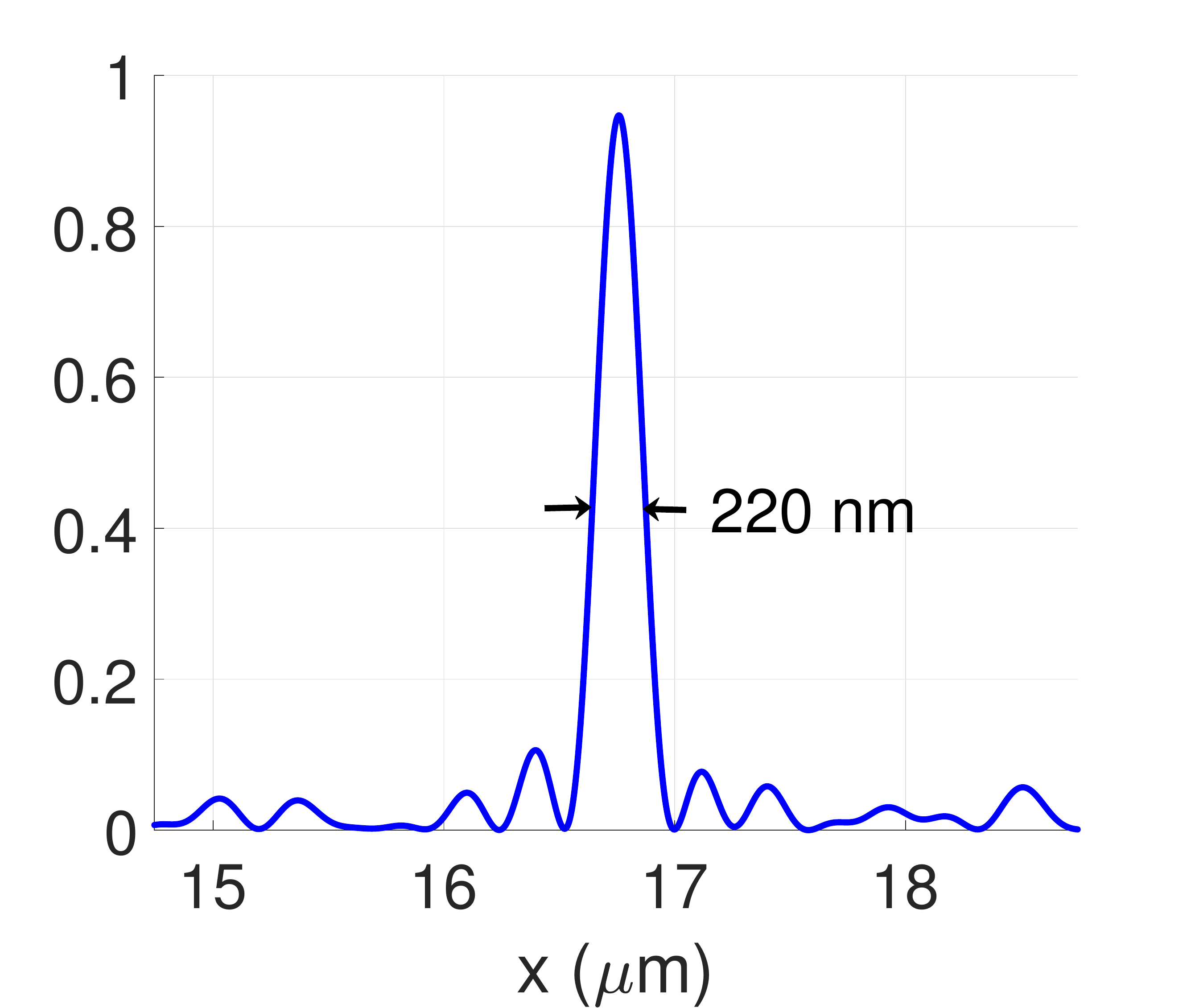}}
        \hspace{1cm}
 \centering
   \subfigure [\label{mod_Ey_long_lamb064_nb103} $|E_y(x_0,z)|^2/\int_{\mathcal{I}}| E_y^{inc}(x,z_s)|^2 dx$]
        {\includegraphics[width=0.35\textwidth]{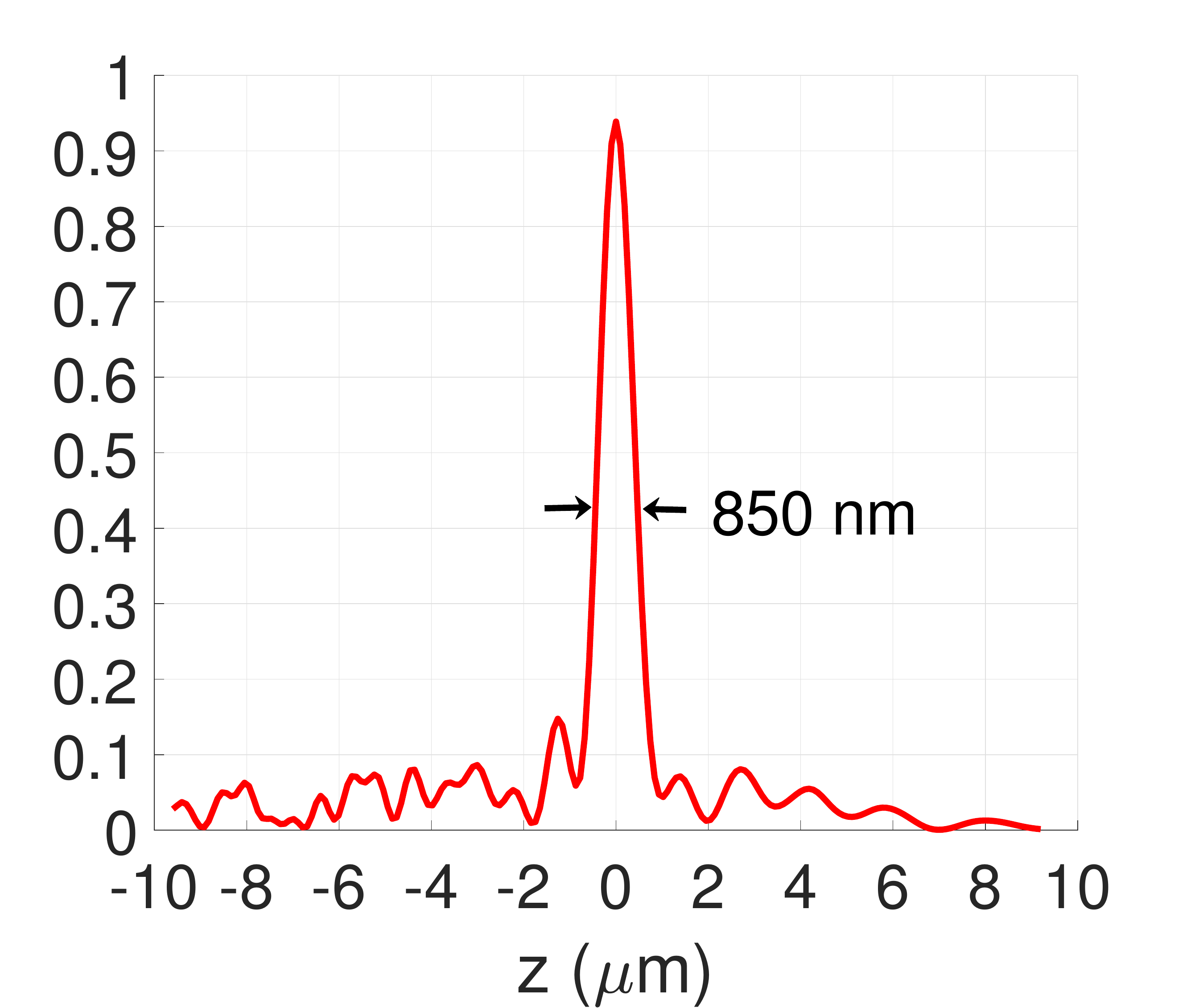}}
 \caption{\label{Ey_long_lamb064_nb103} Design a $52\lambda$ wide metalens consisting of $N_p=203$ nano-rods and air-gaps focusing a normally incident TE-polarized plane wave, with a wavelength $\lambda=0.64\mu m$ at a distance of $15\lambda$: best optimized structure results. Figure \ref{mode_Ey_lambd064_nb103_Px52lambda}: Cartography  of the  normalized  electric field intensity in the plane $(X,O,Y)$. Figure \ref{mode_Ey_lambd064_nb103_Px52lambda}: Cartography  of the  electric field phase in the plane $(X,O,Y)$. Figure \ref{mod_Ey_trans_lamb064_nb103}: line scan of the  normalized  electric field intensity $(Ox)$-axis at  $z_f=0$. Figure \ref{mod_Ey_long_lamb064_nb103}: line scan of the  normalized  electric field intensity along $(Oz)$-axis at $x=x_0$.     Numerical parameters: $\lambda=0.64\mu m$, $N_p=203$, $d=52\lambda$, height $e_z=325 nm$, TE polarization.}
\end{figure}

As a second example, we extend the lens features to a width of $d=52\lambda$ with a pattern comprising $203$ rods/air-gaps.  The  number of Fourier basis functions is also increased in order to ensure the convergence of the numerical results: it is set to $601$ (i.e. $M=300$). 
We report in Figures \ref{mode_Ey_lambd064_nb103_Px52lambda} and \ref{angle_Ey_lambd064_nb103_Px52lambda} the electric field intensity and phase for the best obtained design. A well-defined focusing and wave-front conversion can also be distinguished in this case. Besides, as shown in Figures \ref{mod_Ey_trans_lamb064_nb103} and \ref{mod_Ey_long_lamb064_nb103}, and as expected, both the power energy flow and the resolution of the designed lens are increased when increasing the lens geometrical features. The intensity ratio side-lobe/central-lobe is around $10.5\%$. The LFWHM decreases to $220nm \simeq 0.3438\lambda$ (Fig. \ref{mod_Ey_trans_lamb064_nb103}) while in the axial direction the AFWHM is reduced to $850nm \simeq 1.3281\lambda$ (Fig. \ref{mod_Ey_long_lamb064_nb103}).

Table 1 recaps the computation times required to design a $d$-wide metalens made of $N_p$ nano-rods, with respect to the eigenvalue matrix dimension (conical formulation). The maximum number of iterations is set to 50 iterations.
\begin{table}[h!]
\centering
\begin{tabular}{|c||c c|}
\hline 
{$N_p$} & 52 & 203\\ 
\hline
$d$ & $32\lambda$ & $52\lambda$\\ 
\hline
$dim=2(2M+1)$ &  $2(500+1)$ &   $2(600+1)$\\ 
\hline
computation times (s) &  $4.5\times 10^4s$ &  $2.5\times 10^5s$\\ 
\hline
\end{tabular}
\caption{\label{tab_old_last_mother}  Panel of the computation times for the design of a metalens using the Aperiodic Fourier modal method (AFMM) in its conical formulation, with $2M+1$ Fourier basis functions (the dimension of the eigenvalues equations matrix is then $2(2M +1)\times2(2M +1)$ and the dimension of the scattering matrix using for the longitudinal boundary conditions is $4(2M +1)\times4(2M +1)$ ). 
 }
\end{table}

\begin{figure}[htb!]
\centering
   \subfigure [\label{histo_ek_nb51_Px32} $N_p=52$, $d=32\lambda$]
        {\includegraphics[width=0.4\textwidth]{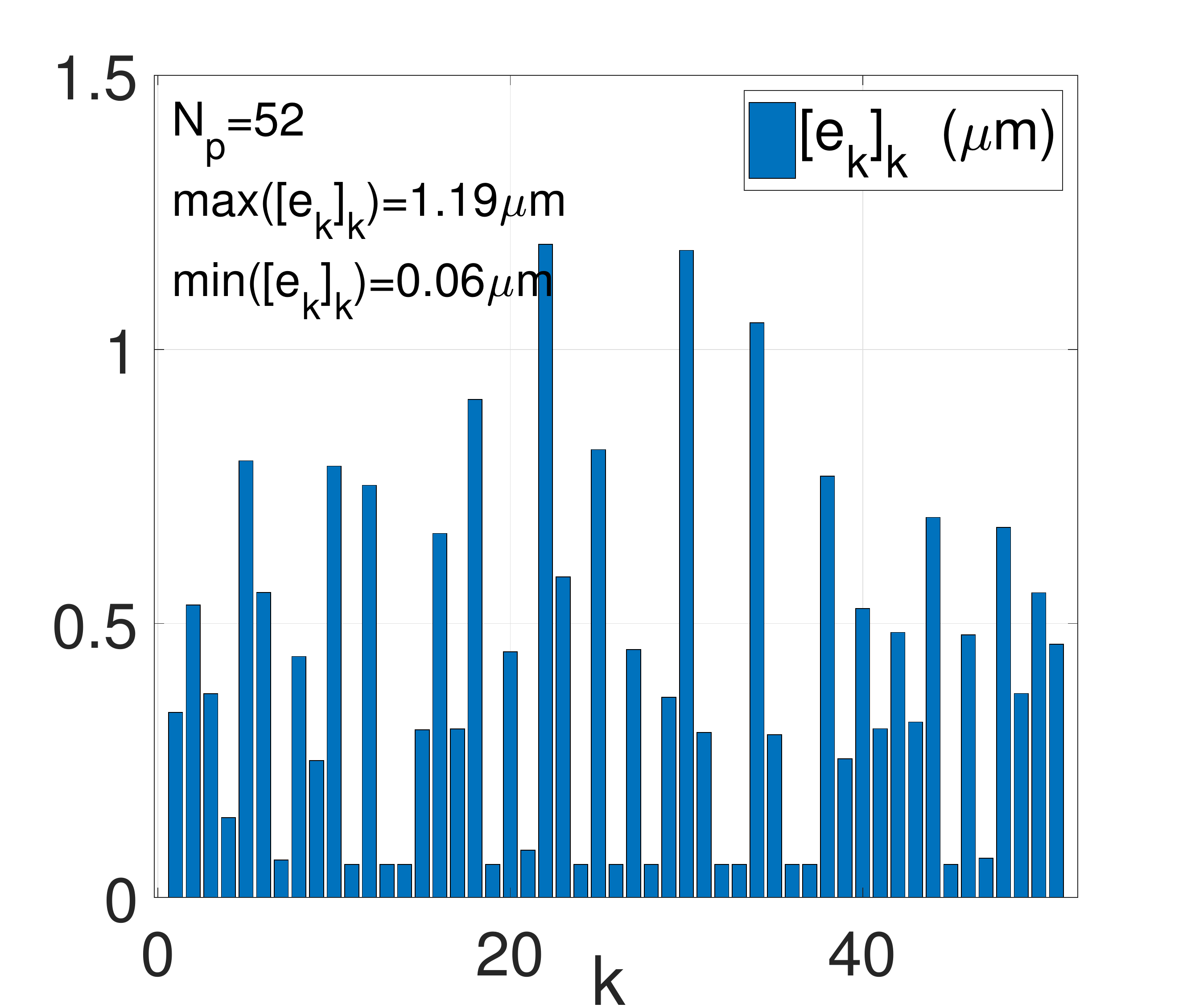}}
 \centering
   \subfigure [\label{histo_ek_nb203_Px52} $N_p=203$, $d=52\lambda$]
        {\includegraphics[width=0.4\textwidth]{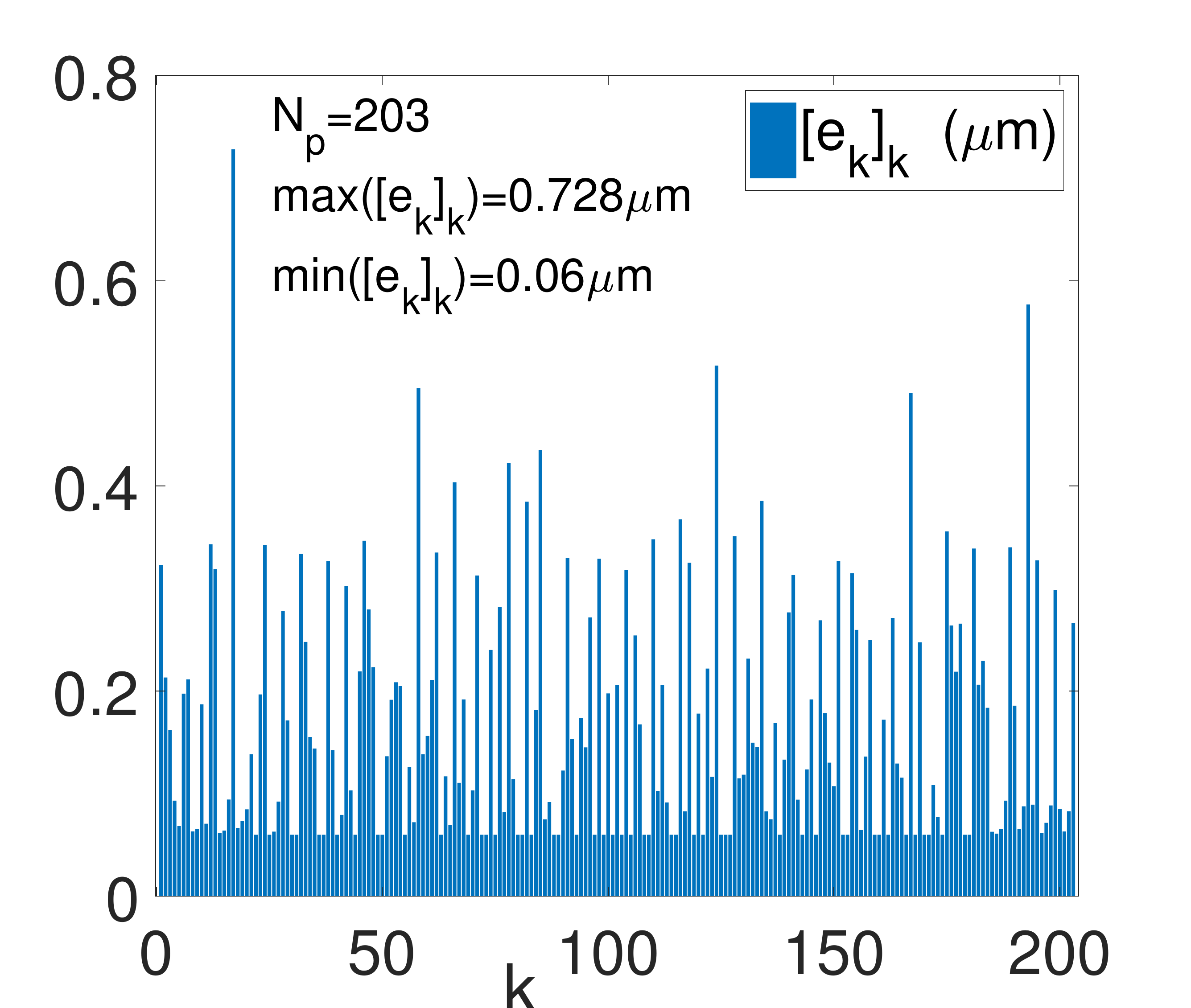}}
 \caption{\label{histo_ek_nb203_Px32_Px52}  Rods/air-gaps widths sequences $[e_k]$ with respect to their locations labelled by $k$.  No symmetry property is apparent in these results. Numerical parameters: $\lambda=0.64\mu m$, $N_p=203$, $d=51\lambda$, $e_z=325 nm$, TE polarization.}
\end{figure}

Finally, it is worth noting that all the optimizations we carried out, systematically yielded original, unexpected, asymmetric on-axis metalenses even though under normal incidence. This is illustrated in Figures \ref{histo_ek_nb51_Px32} and \ref{histo_ek_nb203_Px52} where we report, for the two designed metalenses, the values of the rods/air-gaps) widths sequences $[e_k]$ with respect to their locations labelled by $k$. No symmetry property is apparent in these results indicating that sophisticated energy exchange, between the modes in the design area, takes place and these interactions cannot be explained and cannot be modeled thanks to some trivial rules. The modes contained in  the design area of the structure are not symmetric but their combination yields a symmetric electromagnetic far field.  Without enforcing any  symmetry properties via a curvilinear phase function, for example, the optimization process freely evolves towards unexpected exotic structures.   

\section{Conclusions and outlook}
The inverse design of metasurfaces with the adjoint-based topology optimization method often faces two challenges: the first one is the relatively high number of the design parameters and the second is the efficient numerical resolution of equations handling the physical phenomena of the problem under consideration. In photonics, these equations are the partial differential equations (PDE) obtained from  Maxwell's equations and a set of {\it ad-hoc} boundary conditions.
In this paper, we reported a new concept enabling to overcome the first obstacle. This is accomplished through the introduction of a set of a few, but key, design parameters allowing the fine-tuning of a given metasurface geometrical features. The method has been successfully applied to design a $\lambda$-scale metagrating deflecting a normally incident plane wave into a given direction with high efficiency. The low number of design parameters introduced in our formulation, makes the inverse design much less computational time demanding. We exploited this gain in computational efficiency to perform the design of a large-scale aperiodic metasurface in records times : it consists of a dielectric metalens enabling a high energy flow focusing into a well-defined focus spot. In this case, we optimized the full device in one time, without enforcing any phase conditions (as is usually done for this kind of devices), and have shown that the proposed approach systematically provides original and unexpected asymmetric on-axis metalenses; even though under normal illumination. 

Besides, the optimization of very large-scale aperiodic photonic structures remains an immense computational challenge since one needs a very efficient and fast electromagnetic solver which is the second challenge mentioned above. 
For this key point, \textsl{a priori}, there is no unique and universal method that allows to efficiently solve PDEs obtained from Mawxell's equations in the general case. However, for certain categories of electromagnetic problems with specific geometries and symmetries, it is possible to identify a class of numerical methods allowing to accurately and efficiently solve these PDEs. One-dimensional metasurfaces often consist of an arrangement of rods deposited on a substrate, and Modal Methods (MM) have proven to be very effective in solving such problems. These approaches consist in describing the electromagnetic field in terms of the eigenfunctions of an operator that represents the propagation through the structure. These eigenfunctions are expanded on a set of a given basis of functions and in the particular case of the Fourier Modal Method (used in this work), the Fourier basis functions are used. However, the main drawback of the FMM remains the representation of non smooth fields through finite Fourier sums resulting, inevitably, in slow convergence rates. And the worst cases are those of diffraction problems involving metals or {\bf{very}} large-scale and/or multi-scale structures. These cases often  require a high number of Fourier basis functions to converge which increases the computational times. To remedy this situation, one can replace the FMM by the very efficient and well established Polynomial Modal Method (PMM) \cite{Kofi_gegenbauer1, Kofi_gegenbauer2, Kofi_gegenbauer3, Kofi_2D, Kofi_elsevier} equipped with Perfectly Matched Layers (PMLs); an approach that proves to be more efficient and robust even though at the price of a more complex numerical implementation. 

\section*{Funding Information}
This work has been sponsored by the French government research program "Investissements d'Avenir" through the IDEX-ISITE initiative 16-IDEX-0001 (CAP 20-25)
\bigskip

\section*{Disclosures}
The authors declare no conflicts of interest.



\begin{thebibliography}{999}
\bibitem{Sell1} D. Sell, J. Yang, S. Doshay, R. Yang,  and J.-A. Fan, "Large-angle, multifunctional metagratings based on freeform multimode geometries," Nano Lett. 17, 3752-3757 (2017).

\bibitem{Yang1} J. Yang, and J.-A. Fan,  "Topology-optimized metasurfaces: impact of initial geometric layout," Opt. Letters \textbf{42}, 3161-3164 (2017).  

\bibitem{Sell2} D. Sell, et al., "Ultra-High-Efficiency Anomalous Refraction with Dielectric Metasurfaces," ACS Photonics \textbf{5}, 2402-2407 (2018).

\bibitem{Yang}  J. Yang,  D. Sell,  and J.-A. Fan,  "Freeform metagratings based on complex light scattering dynamics for extreme, high efficiency beam steering," Ann. Phys. \textbf{530}, 1700302 (2018).

\bibitem{EWang}  E. W. Wang, D. Sell, T.  Phan  and J-A. Fan, "Robust design of topology-optimized metasurfaces," Opt. Mater. Express \textbf{9}, 469-482 (2019).


\bibitem{Kofi_TO}  K. Edee,  "Topology optimization of photonics devices: fluctuation-trend analysis concept; random initial conditions with Gaussian and Durden-Vesecky power density bandlimited spectra," JOSAB \textbf{37}, 2111-2120 (2020).

\bibitem{Kofi_TO2}  K. Edee,  et al., "Inverse design of a 1D dielectric metasurface by topology optimization: fluctuations-trend analysis assisted by a diamond-square algorithm," JOSAB \textbf{37}, 3721-3728 (2020).

\bibitem{Shrestha} S. Shrestha, A. C. Overvig, M. Lu, A. Stein, and N. Yu, "Broadband achromatic dielectric metalenses," Light: Sci. Appl. 7(1), 85 (2018).

\bibitem{SWang} S. Wang, P. C. Wu, V.-C. Su, Y.-C. Lai, M.-K. Chen, H. Y. Kuo, B. H. Chen, Y. H. Chen, T.-T. Huang, J.-H. Wang, R.-M. Lin, C.-H. Kuan, T. Li, Z. Wang, S. Zhu, and D. P. Tsai, "A broadband achromatic metalens in the visible,"
Nat. Nanotechnol. 13(3), 227–232 (2018).

\bibitem{Phan} T. Phan, and al. "High-efficiency, large-area, topology-optimized metasurfaces," Light. Sci. Appl. \textbf{8}, 48 (2019).

\bibitem{Chung} H. Chung and O. D. Miller, "High-NA achromatic metalenses by inverse design," Opt. Express 28, 6945-6965 (2020)
\bibitem{Lu}  J. Lu  and J. Vuckovic,  "Nanophotonic computational design," Opt. Express \textbf{21}, 13351-13367, (2013).   

\bibitem{Lalau} Lalau-Keraly, C. M.,  Bhargava, S., Miller, O. D.,  and  Yablonovitch, E. Adjoint shape optimization applied to electromagnetic design. Opt. Express \textbf{21}, 21693-21701 (2013)

\bibitem{Hughes} Hughes, T. W., Minkov, M, Williamson, I. A. D. and Fan, S. Adjoint Method and Inverse Design for Nonlinear Nanophotonic Devices. ACS Photonics 2018, \textbf{5}, 4781-4787, (2018).

\bibitem{Molesky}  Molesky, S. et al., "Inverse design in nanophotonics," Nat. Photon. 12, 659-670, (2018).
\bibitem{Frandsen} L. H. Frandsen et al., "Broadband photonic crystal waveguide $60^o$ bend obtained utilizing topology optimization," Opt. Express 12, 5916-5921 (2004).
\bibitem{Borel}  P. I. Borel,  et al., "Topology optimization and fabrication of photonic crystal structures,"  Opt. Express 12, 1996-2001 (2004).
\bibitem{Piggot} A. Y. Piggot, J. Lu, K. Lagoudakis  et al. "Inverse design and demonstration of a compact and broadband on-chip wavelength demultiplexer," Nat. Photonics 9, 374-377
(2015).
\bibitem{Xiao}  T. P. Xiao, O. S. Cifci, S. Bhargava, H. Chen, T. Gissibl, W. Zhou, H. Giessen, K. C. Toussaint, E. Yablonovitch, and P. V. Braun, "Diffractive spectral-splitting optical element designed by adjoint-based electromagnetic optimization and fabricated by femtosecond 3D direct laser writing," ACS Photonics 3, 886-894 (2016).

\bibitem{Lin}  Z. Lin,  et al., "Topology-optimized multilayered metaoptics," Phys. Rev. Appl. 9, 044030 (2018).














 \bibitem{Knop}  K. Knop, "Rigorous diffraction theory for transmission phase gratings with deep rectangular grooves,"   J. Opt.  Soc.  Am. A {\bf 68}, 1206-1210 (1978)

\bibitem{Granet_Guizal} G. Granet, and B. Guizal,  "Efficient implementation of the coupled-wave method for metallic lamellar gratings in TM polarization,"  J. Opt.  Soc.  Am. A {\bf 13}, 1019-1023 (1996)

\bibitem{Lalanne} P. Lalanne and G. M. Morris, "Highly improved convergence of the coupled-wave method for TM polarization," J. Opt.  Soc.  Am. A  {\bf 13}, 779-784 (1996)

 \bibitem{Li} L. Li,   "Use of Fourier series in the analysis of discontinuous periodic structures," J.  Opt. Soc.  Am. A {\bf 13}, 1870-1876 (1996)
 
 \bibitem{Granet_ASR} G. Granet, "Reformulation of the lamellar grating problem through the concept of adaptive spatial resolution," J. Opt. Soc. Am. A {\bf16} 2510-2516  (1999).

\bibitem{Lalanne_lens} P. Lalanne  and  P. Chavel, "Metalenses at visible wavelengths: past, present, perspectives," Laser \& Photonics Reviews, \textbf{11}, 1600295 (2017)


\bibitem{Yaoyao} Y. Liang, H. Liu, F. Wang, H. Meng, J. Guo, J.  Li, Z. Wei, "High-Efficiency, Near-Diffraction Limited, Dielectric Metasurface Lenses Based on Crystalline Titanium Dioxide at Visible Wavelengths," Nanomaterials \textbf{8},  288 (2018). 


\bibitem{Talukdar}  T. H. Talukdar and  J. D. Ryckman, "Multifunctional focusing and accelerating of light with a simple flat lens,"  Opt. Express \textbf{28},  30597 (2020)
\bibitem{plumey} J.-P. Plumey, K. Edee and G. Granet, "Modal expansion for the 2D Greens function in a non orthogonal coordinate system," Progress In Electromagnetics Research, PIER 59, 101-112. (2006).

\bibitem{Kofi_gegenbauer1} K. Edee, "Modal method based on subsectional Gegenbauer polynomial expansion for lamellar grating,"  J. Opt. Soc. Am.  A {\bf 28}, (2011)

\bibitem{Kofi_gegenbauer2} K. Edee, I. Fenniche, G.  Granet, and B.  Guizal,  "Modal method based on subsectional Gegenbauer polynomial expansion for lamellar gratings: Weighting function, convergence and stability,"  PIER, {\bf 133} 17-35, (2013)

\bibitem{Kofi_gegenbauer3}  K. Edee and B. Guizal,   "Modal method based on subsectional Gegenbauer polynomial expansion for nonperiodic structures: complex coordinates implementation,"  J. Opt. Soc. Am.  A {\bf 30},  4,  631-639 (2013)

\bibitem{Kofi_2D}  K. Edee and  J.-P. Plumey,  "Numerical scheme for the modal method based on subsectional Gegenbauer polynomial expansion: application to biperiodic binary grating," J. Opt. Soc. Am.  {\bf 31}, 402-410 (2015)


\bibitem{Kofi_elsevier} K. Edee,  J.-P. Plumey, and  B. Guizal,"Unified Numerical Formalism of Modal Methods in Computational Electromagnetics and Latest Advances: Applications in Plasmonics," 
Elsevier, Advances in Imaging and Electron Physics Vol. 197, chapter 2  (2016)









\end{thebibliography}






\end{document}